\documentclass[aps,prd,12pt,nofootinbib]{revtex4}
\usepackage{epsfig}
\usepackage{graphicx}
\usepackage[font=small,skip=0pt,justification=raggedright]{caption}
\usepackage{dsfont}
\usepackage{amsmath}
\usepackage{amssymb}
\usepackage{mathrsfs}
\usepackage{verbatim}
\usepackage{dutchcal}

\usepackage[normalem]{ulem}
\usepackage{xcolor}

\newcounter{fig}   \newcommand{\lbfig}[1]{\refstepcounter{fig}
\label{#1} }
\newcommand{\vphi}{\varphi}

\newcommand{\bea}{\begin{eqnarray}}
\newcommand{\eea}{\end{eqnarray}}
\newcommand{\be}{\begin{equation}}
\newcommand{\ee}{\end{equation}}

\newcommand{\re}[1]{(\ref{#1})}

\newcommand{\cgg}{\mathcal{g}}

\newcommand{\ta}{\theta}

\newcommand{\al}{\alpha}

\newcommand{\si}{\sigma}

\newcommand{\pa}{\partial}

\newcommand{\eqn}{\begin{eqnarray}}
\newcommand{\eqnx}{\end{eqnarray}}

\date{\today}

\begin{document}

\title{Boson stars in the $U(1)$ gauged 3+1 dimensional $O(3)$ sigma-model}

\author{A. Mikhaliuk}
\affiliation{Belarusian State University, Minsk 220004, Belarus}
\author{Y. Shnir}%
\affiliation{BLTP, JINR, Dubna 141980, Moscow Region, Russia}

\begin{abstract}

We study regular self-gravitating non-topological soliton solutions of the
$U(1)$ gauged  non-linear $O(3)$ sigma model with the usual kinetic term and a simple symmetry breaking
potential in 3+1 dimensional asymptotically flat 
spacetime. Both parity-even and parity-odd configurations with an angular node of the scalar field are considered. 
Localized solutions are endowed with an electric charge, spinning configurations are also  coupled to the
toroidal flux of magnetic field. 
We confirm that such solutions do not exist  in the flat space limit. Similar to the usual boson stars, a spiral-like frequency dependence of the mass and
the Noether charge of the gauged 
solutions is observed. 
Depending on the
relative strength of gravity and the electromagnetic interaction, the 
resulting gauged $O(3)$ boson stars
at the mass threshold either possess the usual Newtonian limit, or they are linked to a regular strongly gravitating critical configuration.  
We explore domain of existence of the solutions and investigate some of their physical properties.

\end{abstract}
\maketitle

 \section{Introduction } \label{Introduction}

It has been known for a long time that matter fields can be localized by gravity \cite{Wheeler:1955zz}, see e.g. \cite{Jetzer:1991jr,Schunck:2003kk,Liebling:2012fv,Shnir:2022lba} for detailed review. 
So-called {\it boson stars} \cite{Kaup:1968zz,Feinblum:1968nwc,Ruffini:1969qy,Colpi:1986ye}, which represent localized configurations of a self-gravitating oscillating complex scalar field with a harmonic phase, provide a simple example of such configurations. 
Boson stars  have attracted increasing attention in the last decade, in particular as 
black hole mimickers \cite{Guzman:2009zz,Torres:2000dw,Olivares:2018abq} and modeling of galactic halos and their lensing properties \cite{Mielke:2002bp,Schunck:2006rk,Chen:2020cef}.
Further, the boson stars are considered as possible candidates for the hypothetical weakly-interacting ultralight
component of dark matter  
\cite{Preskill:1982cy,Dine:1982ah,Guth:2014hsa,Matos:1998vk,Suarez:2013iw,Levkov:2016rkk}.

 In the simplest case, boson stars arise as solutions of the Einstein-Klein-Gordon theory with a massive complex scalar field coupled to gravity without self-interaction. These mini-boson stars do not have a regular flat spacetime limit. Very large massive boson stars exist in the models with polynomial potentials 
\cite{Kleihaus:2005me,Kleihaus:2007vk} or in the two-component Einstein-Friedberg-Lee-Sirlin model
 \cite{Friedberg:1986tq}. They 
 are linked to the corresponding flat space configurations, which represent stable isospinning non-topological solitons, the Q-balls \cite{Rosen:1968mfz,Friedberg:1976me,Coleman:1985ki}. 
 
Isorotations may also stabilise self-gravitating non-topological solitons in the non-linear $O(3)$ sigma model with a non-negative potential \cite{Verbin:2007fa,Cano:2023bpe,Adam:2025ktm}. There
are families of regular asymptotically flat stationary spinning $O(3)$ sigma model
solutions,
and hairy black holes in a simple model which include only the usual kinetic term and a symmetry breaking potential  
\cite{Verbin:2007fa,Herdeiro:2018djx,Cano:2023bpe}. Dynamical evolutions of such
solutions  closely resemble the pattern found for the spinning complex
scalar fields coupled to gravity, the boson stars and black holes with synchronised hair
\cite{Kaup:1968zz,Feinblum:1968nwc,Ruffini:1969qy,Colpi:1986ye,Hod:2012px,Herdeiro:2014goa}.
These non-topological stationary solutions of the
$O(3)$ non-linear sigma model minimally coupled to gravity trivialize in the flat space limit
\cite{Verbin:2007fa,Herdeiro:2018djx}. Indeed, 
 in 3+1 dimensions, the Derrick’s theorem \cite{Derrick:1964ww} does not allow for the existence
of such static finite energy solitons in the usual $O(3)$ sigma model, which includes only
quadratic in derivative term and a potential.
Such topological solitons, referred as Hopfions, appear in the
scale-invariant Nicole model \cite{Nicole:1978nu} and, more importantly, in the Faddeev-Skyrme model \cite{Faddeev:1976pg},
which contains terms with both two and four derivatives. Apart the addition of higher
derivative terms, no other mechanism to secure stability of the regular topological solitons
of the $O(3)$ sigma model in three spacial dimensions is known. 

Self-gravitating Hopfions in the $O(3)$
Faddeev-Skyrme model are smoothly connected to the corresponding flat space
topological solitons of certain degrees
\cite{Shnir:2015lli}. However, the virial theorem does not exclude existence of regular isospinning $O(3)$ solitons stabilized by internal rotation in the flat space \cite{Herdeiro:2018djx}, certain modification of the potential term may allow for construction of such non-topological solitons in 3+1 dimensional space. 

As it turns out, the boson stars posses a complicated morphology
\cite{Friedberg:1986tq,Yoshida:1997nd,Kleihaus:2007vk,Herdeiro:2019mbz,Cunha:2022tvk,Shnir:2022lba,Liang:2025myf}.
In the simplest case, these solutions are spherically symmetric, also there are
axially symmetric boson stars with a non-zero angular momentum, solutions with negative parity, radially excited
boson stars and a variety of non-trivial multipolar stationary configurations without any continuous symmetries.
Qualitatively, the appearance of such solutions may be related to the force balance between the repulsive scalar
interaction and the gravitational attraction in equilibria. Furthermore,
the local $U(1)$ symmetry of a model supporting Q-balls
can be promoted to the local gauge symmetry, corresponding gauged Q-balls possess electric charge
\cite{Rosen:1968mfz,Lee:1988ag,Anagnostopoulos:2001dh,Gulamov:2013cra,Gulamov:2015fya}.
The presence of the long-range gauge field affects the properties of the solitons, as the gauge
coupling increases, the electromagnetic repulsion may destroy configurations. Gravitational attraction,
on the other hand, may produce an opposite effect, the pattern of dynamical evolution of gauged boson stars can be
rather complicated \cite{Kunz:2021mbm}.

The objective of this paper is to analyze  the properties of the regular self-gravitating solutions of the
$U(1)$ gauged  non-linear $O(3)$ sigma model minimally coupled to
Einstein gravity, focusing our study on non-topological localized configurations
of different types, and determine their domains of existence.

This paper is organized as follows. In Section II, we introduce the model,
and the field equations with the stress-energy tensor of the system of interacting fields.
Here we describe the axially-symmetric parametrization of the metric and the matter fields and
the boundary conditions imposed on the configuration.
We also discuss the physical quantities of interest. 

In Sec. III we present the
results of our study of self-gravitating solitons with particular emphasis on the role of the electromagnetic interaction.
We conclude with a discussion of the results and final remarks.

\section{The model}
\subsection{The action and the field equations}
We consider the $U(1)$ gauged non-linear sigma-model,
which is minimally coupled to Einstein's gravity in a
(3+1)-dimensional space-time. The corresponding action
of the system is
\be
S=\int d^4 x \sqrt{-{\mathcal g}} \left(\frac{R}{16\pi G} - L_{m}\right),
\label{lag}
\ee
where the rescaled matter field Lagrangian is
\be
L_{m} = -\frac14 F_{\mu\nu}F^{\mu\nu} + \frac12 (D_\mu\phi^a)^2 + U(\phi) \, .
\label{Gauged_O3}
\ee
Here $R$ is the Ricci scalar associated with the spacetime metric $g_{\mu\nu}$ with the determinant
${\mathcal g}$ and $G$ is the Newton's constant.
In the matter field Lagrangian the real triplet of the scalar fields $\phi^a$, $a=1,2,3$, is restricted to
the surface of the
unit sphere, $(\phi^a)^2=1$ and the symmetry breaking potential is 
\be
U(\phi)=\mu^2 (1 -\phi^3)\, ,
\label{U}
\ee
where a mass parameter $\mu$ is a constant\footnote{There are other choices of the potential, see \cite{Verbin:2007fa,Cano:2023bpe,Adam:2025ktm}.  }.

Such potential breaks $O(3)$ symmetry to the subgroup $SO(2)$, and the corresponding Noether current can be written as follows:
\be
j^\mu=-\phi^1 D^\mu\phi^2 + \phi^2 D^\mu\phi^1
\label{ncurrent}
\ee

The $U(1)$ field strength tensor is
$F_{\mu\nu}=\partial_\mu A_\nu-\partial_\nu A_\mu$,
and the covariant derivative of the  field $\phi^a$, that minimally
couples the scalar field to the gauge potential is
\be
D_\mu \phi^\alpha = \partial_\mu \phi^\alpha
+e\,A_\mu\,\varepsilon_{\alpha\beta}\,\phi^\beta \,, \qquad D_\mu
\phi^3 = \partial_\mu \phi^3, \qquad \alpha,\,\beta=1,\,2 \,
\label{covariant}
\ee
with the gauge coupling  $e$. The vacuum boundary conditions are
$\phi_{\infty}^a = (0,0,1),\, D_\mu\phi^a =0,\, F_{\mu\nu}=0$. In the static gauge, where the fields have no explicit dependence on time, the  asymptotic boundary conditions on the gauge potential are 
\be
A_0(\infty)=V,\quad A_i(\infty)=0
\ee
where $V$ is a real constant. 
The model \re{Gauged_O3} is invariant with respect to
the local Abelian gauge transformations
\be
(\phi^1 + i \phi^2) \to
e^{ie \zeta}(\phi^1 + i \phi^2), \quad A_\mu \to A_\mu - \partial_\mu \zeta  
\label{rotate}
\ee
where $\zeta$ is a real function of the coordinates. We remark that asymptotic value of the electric potential $A_0(\infty)$
can be adjusted via the residual $U(1)$ transformations, choosing  $\zeta =-Vt$. In the stationary gauge $A_0(\infty)=0$ and two components \re{rotate} of the scalar triplet acquire an explicit harmonic time  dependence with frequency $\omega=eV$.

Variation of the action (\ref{lag})
with respect to metric yields the Einstein equations:
\be
R_{\mu\nu} -\frac12 R g_{\mu\nu} = 2\alpha^2 \left( T_{\mu\nu}^{Em} + T_{\mu\nu}^{\phi}\right)
\label{einsteq}
\ee
where  $\alpha^2=16\pi G$ is
the effective gravitational coupling and the electromagnetic and scalar components of the energy-momentum tensor are
\be
\begin{split}
T_{\mu\nu}^{Em} &=F_\mu^\rho F_{\nu\rho} - \frac14 g_{\mu\nu} F_{\rho\sigma}  F^{\rho\sigma}\,,\\
T_{\mu\nu}^{\phi} &= D_\mu\phi^a D_\nu\phi^a
 - g_{\mu\nu}\left[\frac{g^{\rho\sigma}}{2}  D_\rho\phi^a
D_\sigma\phi^a +  U(\phi)\right] \, .
\label{Teng}
\end{split}
\ee
We notice that the sigma model constraint
does not allow us to absorb the mass  parameter $\mu$ into the scaled expression for the scalar field.

We obtain field equations by variation of the action (\ref{lag}) with the respect to the field $A_\mu$ and $\phi_a$:
\be
    \partial_\mu\left(\sqrt{-{\mathcal g}}  F^{\mu\nu}\right)=e \sqrt{-{\mathcal g}} j^\nu, \quad D^\mu D_\mu \phi^a + \phi^a( D^\mu \phi^b \cdot D_\mu \phi^b)+ \mu^2\left[ \phi^a(\phi^b\cdot \phi^b_\infty) -\phi^a_\infty\right]=0\, ,
\label{eqfield}
\ee
where we take into account dynamical constraint imposed on the $O(3)$ field. 

The asymptotic expansion of the fields around the vacuum  
\be
\phi^a \approx \phi^a_\infty + \delta \phi^a, \qquad A_\mu\approx a_\mu + \delta_{0\mu} (eV -\omega)  
\ee
where $\delta \phi^a =(1-v_0, \, v_1,\, v_2)$, 
gives the effective stationary Lagrangian of linearized scalar excitations
\be
L_{pert} = \frac12 (\partial_i v_\alpha)^2 +\mu^2 v_\alpha^2 -(eV -\omega)^2 v_\alpha^2  
\ee
Here $\alpha=1,2$ and we take into account the residual gauge invariance of the system. Note that the fluctuations of the component $v_0$ correspond to the second order
of the expansion around the vacuum. Hence, 
the squared effective  mass of the scalar excitations 
$v_\alpha$ is $m_{eff}^2=\mu^2 - (eV -\omega)^2$ and
localized massive scalar modes with exponentially decaying tail may exist if $m_{eff}^2$ is positive. In the stationary gauge $V=0$ and the 
self-gravitating non-topological $O(3)$ solitons 
may occur within a restricted interval
of values of the angular frequency
$\omega\in [\omega_{min},  \mu].$ The configurations smoothly
approach perturbations around Minkowski spacetime and delocalize as $\omega \to \mu$. 
This is usually
referred to as the Newtonian limit.

\subsection{Self-gravitating solitons: The ansatz}
We are interested in stationary axially-symmetric regular localized solutions of the system of equations (\ref{einsteq}) and (\ref{eqfield}), which are similar to the usual boson stars. We adopt the isotropic Lewis-Papapetrou-type 
metric in adapted “quasi-isotropic” spherical coordinates $(t; r; \theta; \varphi)$, for which the line element reads

\be
    ds^2=-e^{f_0(r,\ta)}dt^2+e^{f_1(r,\ta)}(dr^2+r^2d\ta^2)+e^{f_2(r,\ta)}r^2\sin^2{\ta}(d\vphi-\frac{W(r,\ta)}{r}dt)^2 \, .
    \label{ansmetr}
\ee
The
Minkowski spacetime asymptotic is approached for $r\to \infty$ with $f_0=f_1=f_2=W =0$. 
Spherically symmetric solutions correspond to the reduction $W(r,\ta)=0$, $f_1(r)=f_2(r)$, both  metric functions $f_0$ and $f_1$ then depend on the isotropic radial coordinate $r$ only.

For the scalar $O(3)$ field, we adopt a stationary ansatz with a
harmonic time dependence, which is  similar  to the corresponding  parametrization of the  $O(4)$ Skyrme model \cite{Battye:2014qva,Ioannidou:2006nn,Perapechka:2017bsb,Herdeiro:2018daq} and the $U(1)$ gauged Skyrmions \cite{Kirichenkov:2023omy,Livramento:2023keg,Livramento:2023tmm}, 
\be
    \phi^a=[\sin{f(r,\ta)}\cos{(n\vphi-\omega t)},\text{ }\sin{f(r,\ta)}\sin{(n\vphi-\omega t)},\text{ }\cos{f(r,\ta)}].
    \label{anssfield}
\ee

The Ansatz for the $U(1)$ potential 
contains two real functions – an electric and a magnetic potential:
\be
    A_\mu dx^\mu=A_t(r,\ta)dt + A_\vphi(r,\ta)\sin{\ta}d\vphi
    \label{ansempot}
\ee
Recall that in the  stationary gauge    $A_t=A_\vphi=0$ at spacial infinity. 

The $U(1)$ gauged charged self-gravitating solutions are characterized by a set of physical observables, the Arnowitt-Deser-Misner (ADM) mass $M$, the angular momentum $J$, the electric charge $Q_e$ and the dipole moment $\mu_m$. 
These quantities can be extracted from the asymptotic behaviour of the metric and the gauge field functions, as
\be
\begin{split}
g_{tt}&\rightarrow -1+\frac{\alpha^2 M}{\pi r}+O(\frac{1}{r^2}),\quad g_{t\phi}\rightarrow  -\frac{\alpha^2 J}{\pi r}\sin^2{\ta}+O(\frac{1}{r^2})\\
A_t & \rightarrow \frac{Q_e}{r} +O(\frac{1}{r^2}),\quad A_\vphi \rightarrow \frac{\mu_m \sin^2\theta}{r^2} +O(\frac{1}{r^3})   
\end{split}
\ee

The ADM mass and the total angular momentum can also be computed as the Komar integrals of the corresponding densities of the total  stress-energy tensor $T_{\mu\nu}=T^\phi_{\mu\nu} + T^{Em}_{\mu\nu}$, 
\be
M=\int\sqrt{-\cgg} dr d\theta d\varphi\left( T^\mu_\mu - 2 T^t_t\right),\quad J=\int\sqrt{-\cgg} dr d\theta d\varphi T^t_\varphi
\label{mass-mom}
\ee
In other words, the mass $M$ and the angular momentum $J$ are conserved charges associated with the two 
Killing vectors, $\partial_t$ and $\partial_\vphi$, respectively.

The temporal component of the current density \re{ncurrent} yields the Noether charge
\be
    Q=\int \sqrt{-\cgg} dr d\theta d\varphi ~j^0=2\pi\int_0^\infty dr\int_0^\pi d\ta\sin^2{f}\left(\omega-\frac{nW}{r}\right)e^{f_1+\frac12 f_2 - \frac12 f_0}r^2\sin{\ta}.
\ee
where we make use of the parametrization of the metric (\ref{ansmetr}) and the $O(3)$ field (\ref{anssfield}). One can show that $J$, $Q$ and the electric charge  $Q_e$ are proportional, see e.g.  ref.  \cite{Schunck:1996he,Collodel:2019ohy,Herdeiro:2021jgc}:
\be
J = nQ =\frac{nQ_e}{e} 
 \label{Jn}
\ee
where $n$ is the winding number of the
scalar field. 
We also note that we consider regular self-gravitating configurations with zero entropy and without an intrinsic temperature. However, they obey the truncated first law of thermodynamics without the entropy term:
$dM=\omega dQ$, see e.g.  
\cite{Herdeiro:2021mol,Stotyn:2013spa}.

Substitution of the ansatz \re{ansmetr},\re{anssfield} and \re{ansempot} into general field equations \re{einsteq}, \re{eqfield} reduces them to a system of seven coupled nonlinear partial differential equations in $r$ and $\theta$. In order to find asymptotically 
flat and regular solutions, appropriate boundary conditions need to be imposed. 
The assumption of asymptotic flatness at the spacial infinity with our choice of the stationary gauge for the Maxwell potentials, gives
\be
    f\vert_{r\rightarrow{\infty}}=f_0\vert_{r\rightarrow{\infty}}=
    f_1\vert_{r\rightarrow{\infty}}=f_2\vert_{r\rightarrow{\infty}}=
    W\vert_{r\rightarrow{\infty}}=A_t\vert_{r\rightarrow{\infty}}=
    A_\vphi\vert_{r\rightarrow{\infty}}=0
    \label{boundradinf}
\ee
On the symmetry axis, the regularity and the axial symmetry implies the following conditions ($n\neq 0$)
\be
\begin{split}
    \pa_{\ta}W\vert_{\ta=0,\pi}=\pa_{\ta}f_0\vert_{\ta=0,\pi}=
    \pa_{\ta}f_1\vert_{\ta=0,\pi}=\pa_{\ta}f_2\vert_{\ta=0,\pi}=0\,,\\
    f\vert_{\ta=0,\pi}=
    \pa_{\ta}A_\vphi\vert_{\ta=0,\pi}=
    \pa_{\ta}A_t\vert_{\ta=0,\pi}=0\,.
\end{split}
    \label{boundang1}
\ee
For the spherically symmetric configuration ($n=0$) we impose $\pa_{\ta}f\vert_{\ta=0,\pi}=0$. 

Further, the condition of the absence of a conical singularity at the symmetry axis
requires that the deficit angle should vanish, $\delta=2\pi\left(1-\lim_{\ta\to 0}\frac{f_2}{f_1}\right)=0$. Therefore, the solutions should satisfy the constrain $f_1\vert_{\ta=0,\pi}=f_2\vert_{\ta=0,\pi}=0$, see  \cite{Herdeiro:2018djx,Herdeiro:2021mol}.
In our numerical scheme we explicitly
imposed this condition.

We assume the solutions are $\mathbb{Z}_2$-symmetric under reflections with respect to the equatorial plane. Thus,  
there are two different types of axially symmetric regular solutions, which possess different parity \cite{Kleihaus:2005me,Kleihaus:2007vk}. 
For  a system with even parity ($\mathcal{P}=1$), all angular derivatives must vanish on the equatorial plane
\be
\begin{split}
    \pa_{\ta}W\vert_{\ta=\frac{\pi}{2}}=\pa_{\ta}f_0\vert_{\ta=\frac{\pi}{2}}=
    \pa_{\ta}f_1\vert_{\ta=\frac{\pi}{2}}=0\,,\\
    \pa_{\ta}f_2\vert_{\ta=\frac{\pi}{2}}=
    \pa_{\ta}A_\vphi\vert_{\ta=\frac{\pi}{2}}=
    \pa_{\ta}A_t\vert_{\ta=\frac{\pi}{2}}=0\,.
\end{split}
    \label{boundang2}
\ee
and $\pa_{\ta}f\vert_{\ta=\frac{\pi}{2}}=0$. For the parity-odd  ($\mathcal{P}=-1$) solutions the boundary condition on the $O(3)$ field is different, $f\vert_{\ta=\frac{\pi}{2}}=0$. 

Finally, regularity of the solutions at the origin requires that both for the parity-even and parity-odd axially-symmetric configurations ($n\neq 0$) we impose
\be
    \begin{split}
    f\vert_{r=0}= W\vert_{r=0}=\pa_{r}f_0\vert_{r=0}=
    \pa_{r}f_1\vert_{r=0}=0\,,\\
    \pa_{r}f_2\vert_{r=0}=
    A_\vphi\vert_{r=0}=
    \pa_{r}A_t\vert_{r=0}=0\,.
\end{split}
    \label{boundrad0}
\ee
For the spherically symmetric solution ($n=0$) the boundary condition on the scalar field is different, $\pa_{r}f\vert_{r=0}=0 $.

\section{Results}
\subsection{Numerical approach}
We have solved the boundary value problem subject to the boundary conditions (\ref{boundradinf})-(\ref{boundrad0}) above with a forth-order finite difference scheme, where the system of seven differential equations is discretized on a grid with a typical size of $200\times 30$
points. In our numerical calculations we introduce a
new compact radial coordinate 
\begin{equation}
	x = \frac{r}{c+r}\, , 
\label{comp_coord}
\end{equation}
which maps the semi-infinite region $[0;\infty)$ onto the unit interval $[0; 1]$. Here $c$ is an arbitrary constant which is used to adjust the contraction of
the grid. In our calculations, we typically take $c\in [0.1,20]$.

The corresponding system of nonlinear algebraic equations
has been solved using the Newton-Raphson
scheme with the Intel MKL PARDISO sparse direct solver \cite{MKL}. Calculations have been performed with the package equipped with an adaptive mesh selection procedure, and the CESDSOL library\footnote{Complex Equations-Simple Domain partial differential equations SOLver, a C++ package developed by I.~Perapechka, see, e.g., Refs. \cite{Kunz:2019sgn,Herdeiro:2019mbz,Herdeiro:2021jgc}}. Typical errors are of order of $10^{-8}$. The solutions do not exhibit conical singularities at the level
of numerical accuracy, numerical calculations confirm that both the Ricci and the Kretschmann
scalars are finite.

The morphology of so constructed solutions can be very complicated \cite{Kleihaus:2005me,Kleihaus:2007vk,Yoshida:1997nd,Herdeiro:2021mol,Herdeiro:2020kvf}. In Fig.~\ref{fig1} we displayed, as an example, four different types of the gravitating soliton solutions of the $O(3)$ sigma model to be discussed below. The parity-even configurations possess spherical or axial symmetry while the parity-odd solutions represent multi-nodal configurations which are  symmetrically aligned along the symmetry axis. 

\subsection{Spherical solutions}
Assuming the spherical symmetry ($n=0$),   the system of field equations \re{einsteq}, \re{eqfield} is reduced to the system of four coupled ordinary differential equations of functions $f,f_0,f_1$ and $A_t$ are functions of $r$ only. Further, in such a case we can 
employ Schwarzschild-like coordinates, with a line element
\be
ds^2=-N(r)\sigma^2(r)dt^2 + \frac{1}{N(r)}dr^2 + r^2(d\theta^2 + \sin^2 \theta d\varphi^2)
\ee
where $N(r)=1-\frac{2m(r)}{r}$. The ADM mass of the solution is determined by the asymptotic value of the mass function $m(\infty)$. 

With this parametrization the
reduced curvature scalar is
$$
R=-2\sigma(N-1-rN^\prime)
$$
and problem reduces to solving two second order
equation for the scalar profile function $f$ and electric potential $A_t$ (in the static gauge), and two first order equations for the metric functions $m$ and $\sigma$:
\begin{equation}
    (r^2N\sigma f')'=r^2\sigma\left(\mu^2-\frac{e^2A_t^2}{N\sigma^2}\cos{f}\right)\sin{f},
\label{f-eqs}    
\end{equation}
\begin{equation}
    m'=\frac{\al^2r^2}{8}\left(Nf'^2+4\mu^2\sin^2{\left(\frac{f}{2}\right)}+\frac{e^2A_t^2}{N\sigma^2}-\frac{A_t'^2}{\sigma^2}\right),
\end{equation}
\begin{equation}
\sigma'=\frac{\al^2r\sigma}{4}\left(f'^2+\frac{e^2A_t^2}{N^2\sigma^2}\sin^2{f}\right),
\end{equation}
\begin{equation}
    (r^2\si A_t')'=2r^2A_t'\si'-\frac{er^2A_t}{N}\sin^2{f}.
\label{At-eqs}    
\end{equation}

\begin{figure}[t!]
\begin{center}
\includegraphics[height=.19\textheight,  angle =-0]{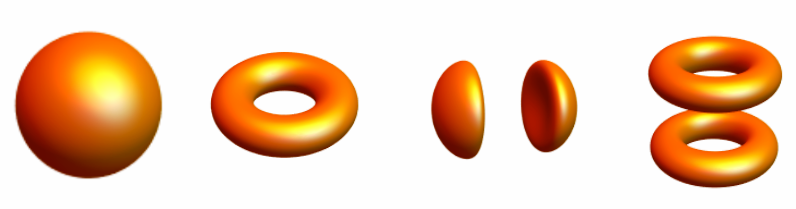}
\end{center}
\caption{\small 
Illustrative examples of different types of solitonic solutions of the $O(3)$  sigma model. The surfaces of constant charge density for fundamental spherically symmetric solution, axially symmetric $n=1$ solution, dipole $n=0$ solution and parity-odd $n=1$ solution, from left to right, respectively, are plotted for  gravitational coupling $\alpha^2=1$, frequency $\omega=0.85$, and gauge coupling $e=0.2$.
}
    \lbfig{fig1}
\end{figure}

The system of equation \re{f-eqs}-\re{At-eqs} can be solved numerically
subject to the boundary conditions 
\be
\begin{split}
    N\vert_{r=0}&=1,\quad \pa_{r}A_t\vert_{r=0}=
\pa_{r}\sigma\vert_{r=0}=
\pa_{r}f\vert_{r=0}=0,\\
N\vert_{r=\infty}&=1,\quad A_t\vert_{r=\infty}=0,~~
\sigma\vert_{r=\infty}=1,~~
f\vert_{r=\infty}=0
\end{split}
\ee
The Schwarzschild-like parametrization 
is useful to analyze possible formation of the event horizon and asymptotic behavior of the solutions, cf. corresponding equations for regular ungauged self-gravitating solutions of the $O(3)$-sigma model \cite{Herdeiro:2018djx}. On the other hand, 
it also can be studied in isotropic coordinates, which is more convenient  for our purposes, comparing various solutions with spherical and axial symmetry. 

First, we can choose $\alpha^2=1$, $\mu^2=1$ and make use of the the time dependent gauge.  
This allows us to compare our results with those obtained in \cite{Herdeiro:2018djx} for ungauged gravitating $O(3)$ solutions. There are two continuous
input parameters, the frequency $\omega$, the gauge coupling $e$ and one discrete parameter $n$.

The spherically symmetric solutions are fundamental states, for which the matter fields  profiles possess no nodes both in radial and angular directions. As the gauge coupling $e$ is set to be zero, regular gravitating boson-star-like solutions of the non-linear $O(3)$ sigma model are recovered \cite{Herdeiro:2018djx}. 
A family of charged solutions  emerge from the uncharged regular boson-star-like solutions when the parameter $e$ is gradually increased from zero while the other input parameters remain fixed.

\begin{figure}[t!]
\begin{center}
\includegraphics[height=.28\textheight,  angle =-0]{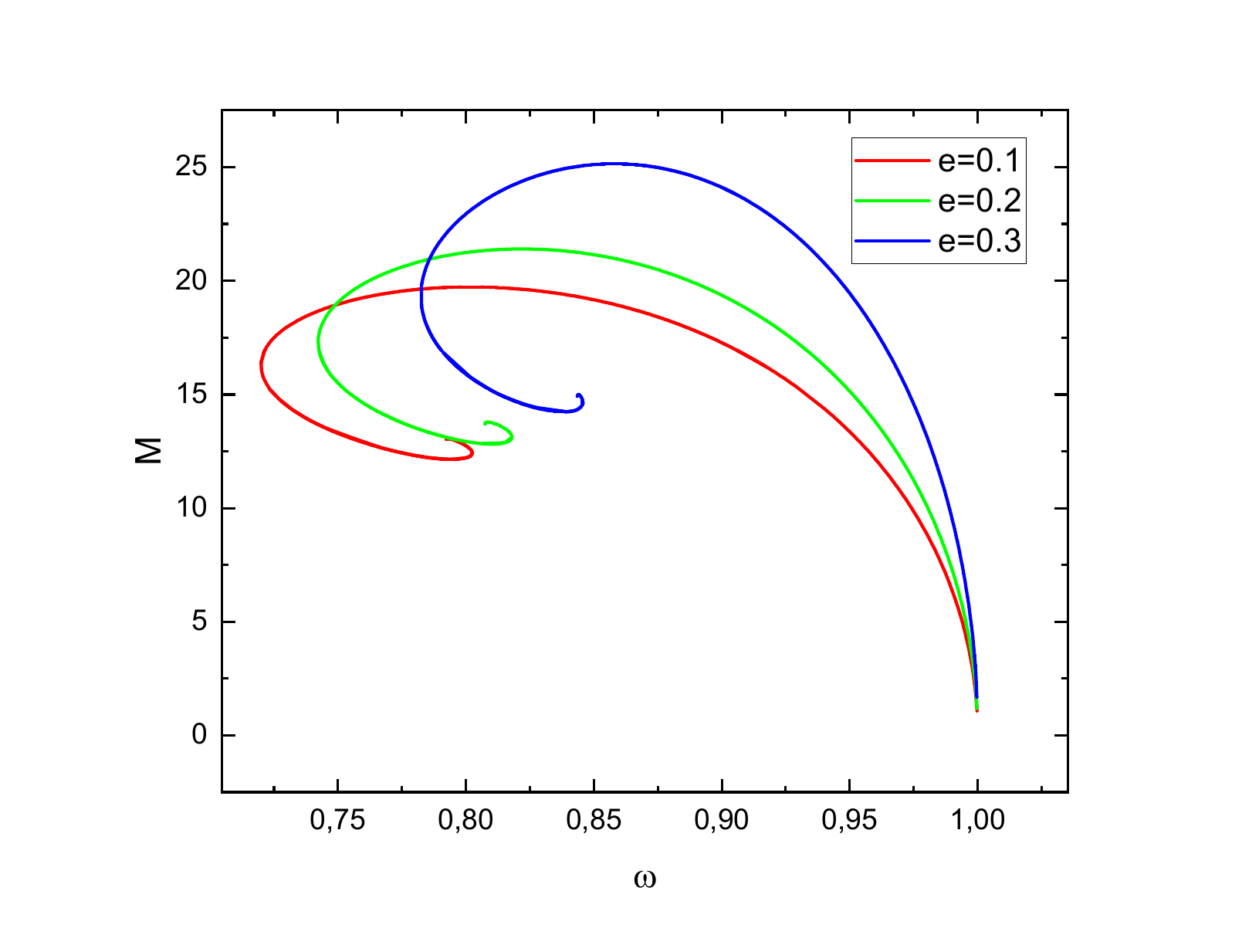}
\includegraphics[height=.28\textheight,  angle =-0]{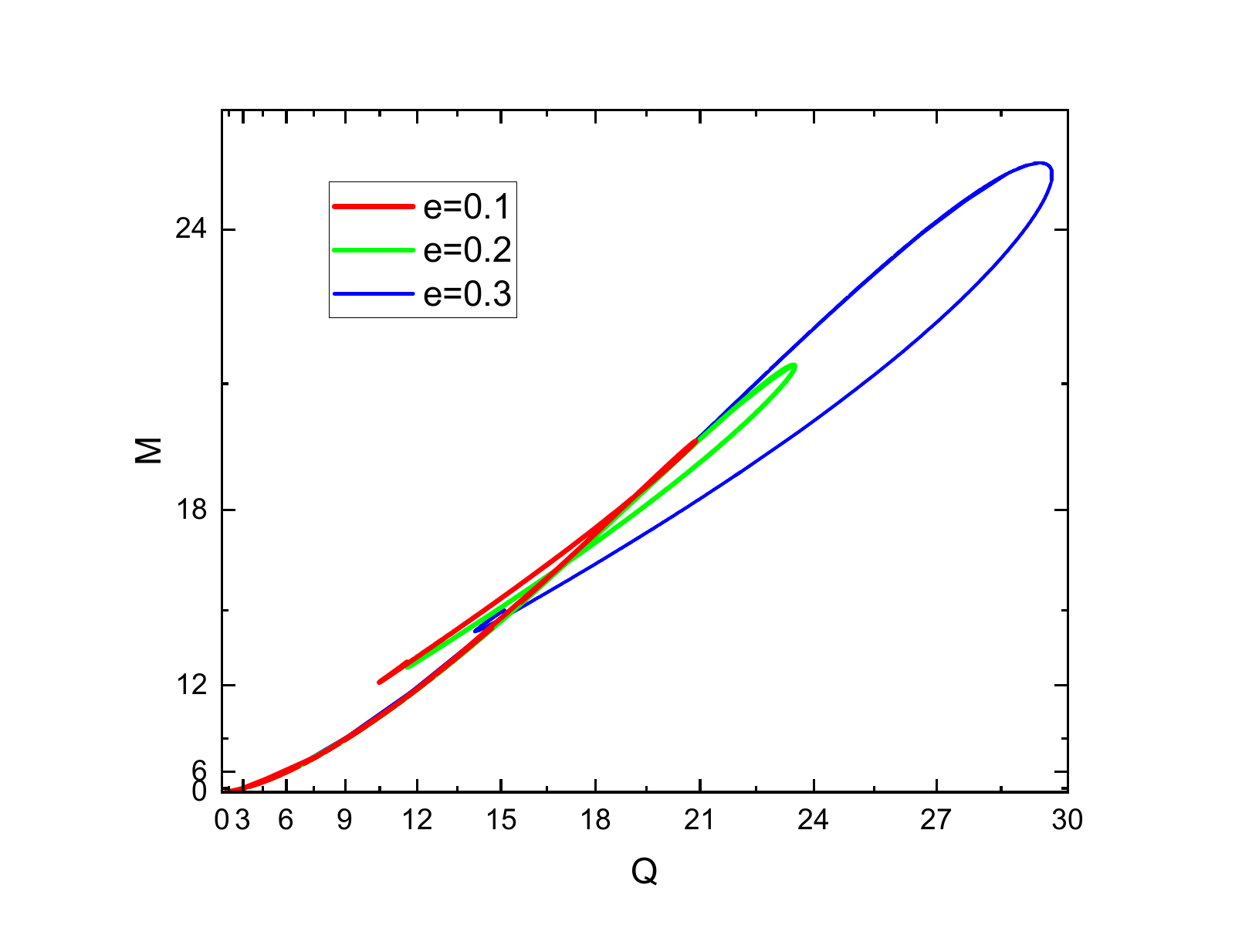}
\includegraphics[height=.28\textheight,  angle =-0]{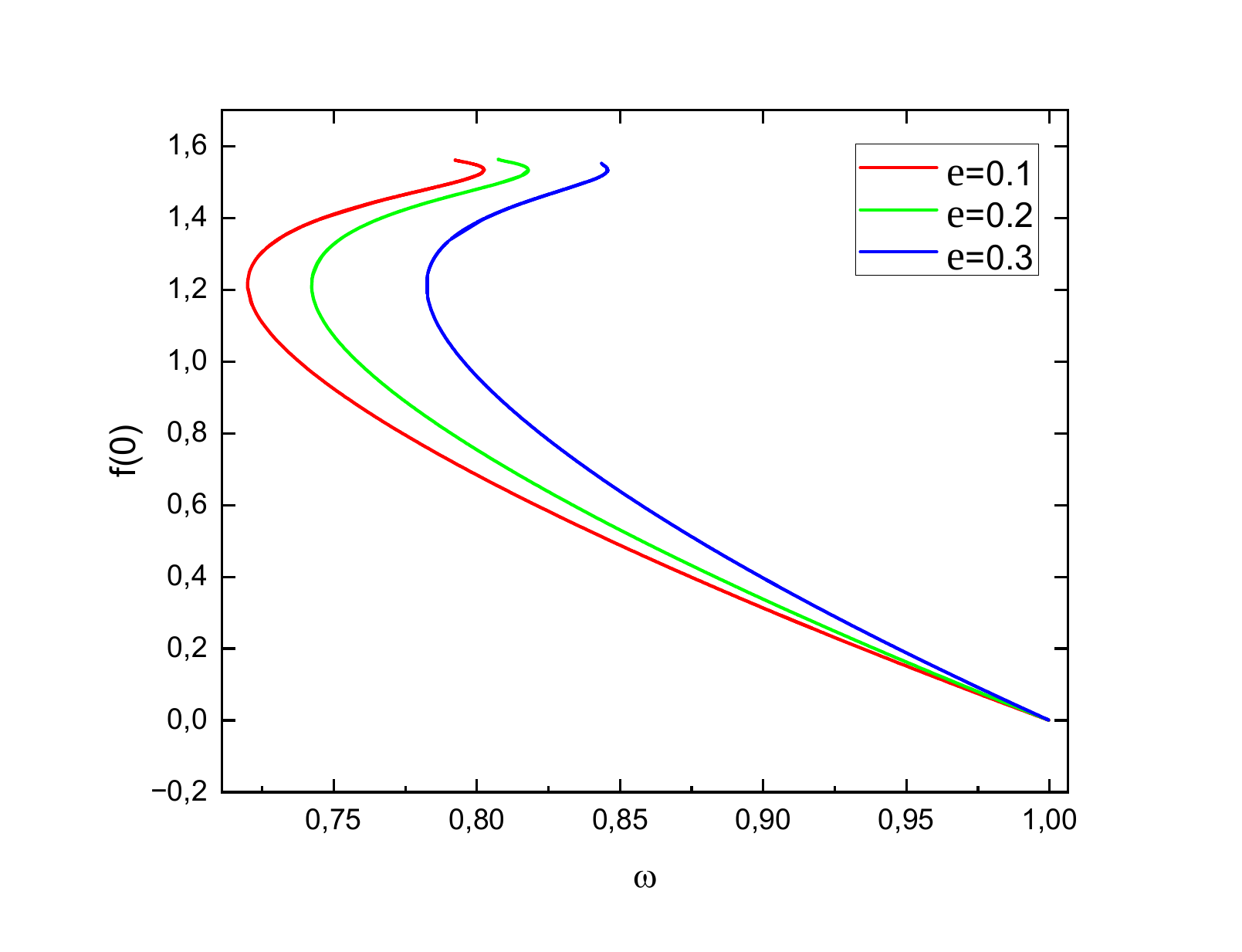}
\includegraphics[height=.28\textheight,  angle =-0]{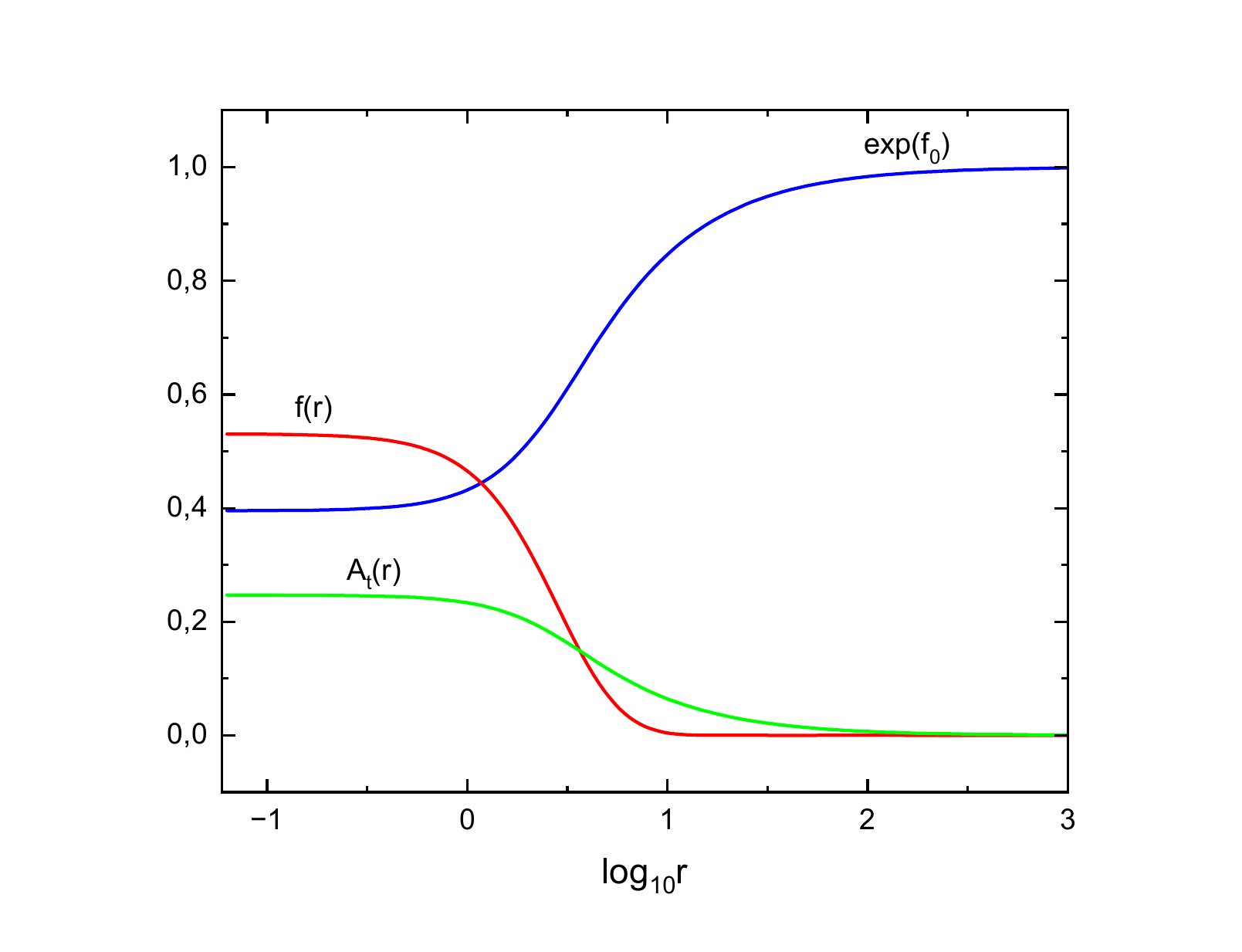}
\end{center}
\caption{\small Spherically symmetric $U(1)$ gauged gravitating $O(3)$ solitons: The ADM mass (upper left)  
and the maximal value of the profile function $f(0)$
(bottom left) vs. frequency $\omega$  and the mass vs. Noether charge (upper right) are plotted for  some set of values of the gauge coupling $e$ and for gravitational coupling $\alpha^2=1$. The profile functions of an illustrative solution on the first branch are plotted for  gravitational coupling $\alpha^2=1$, frequency $\omega=0.85$, and gauge coupling $e=0.2$ (bottom right). 
}    \lbfig{fig2}
\end{figure}

The basic properties of the $U(1)$ gauged gravitating $O(3)$ solutions so constructed can be summarized as follows:

The configurations possesses a core in which most energy is localized, and an asymptotic tail, which is responsible for
the long-range interaction between the well-separated solitons. 
Similar to the usual boson stars, for a given value of the gauge coupling $e$ regular solutions exist for a limited range of frequencies $\omega\in [\omega_{min}, 1]$, where upper critical value corresponds to the mass threshold.
The domain of existence of such $O(3)$ boson stars, in an ADM mass $M$ vs. angular frequency diagram is presented in Fig.~\ref{fig2}, upper left plot.

The minimal critical frequency increases with $e$, our numerical results suggest that  there is
no branch of charged  gravitating $O(3)$ solitons linked to the flat space limit in the model with the potential \re{U}.  

The frequency dependence of the solutions is qualitatively similar to that found for the charged boson stars, see \cite{Lee:1991ax,Jetzer:1989av,Jetzer:1989us,Collodel:2019ohy,Pugliese:2013gsa}. As the angular frequency is decreased below the upper critical value $\omega_{max}=1$, a fundamental branch of charged self-gravitating solutions arises from the trivial scalar vacuum $\phi^a=(0,0,1)$.  
This branch extends backward up to a minimal
value of the frequency $\omega_{min}$, at which it bifurcates with the second forward branch, see Fig.~\ref{fig2}, upper left plot. Both the mass and the charge of the configuration reach the maximum values on the first branch at some critical frequency. The maximal
value of the mass and the charge 
increases with $e$. 

Subsequently, further bifurcations follow and a 
multi-branch structure arises providing an overall inspiralling pattern 
for the  frequency dependency of both charge and mass of the  fundamental $U(1)$ gauged  gravitating $O(3)$ solitons. This behavior is typical for the usual boson stars, see \cite{Lee:1991ax,Jetzer:1989av,Jetzer:1989us}. Both the mass and the charge of the configurations tend to some finite limiting values at the centers of the spirals. When following the spirals towards their centers, the maximal value of the field amplitude $f$ is increasing monotonically, as displayed in Fig.~\ref{fig2}, bottom left plot. Similar to the case of the usual boson stars, the metric function $g_{00}$ at the center of the configuration approaches zero in this limit. 
However, the central value of the scalar amplitude cannot diverge because of the constraint of the non-linear $O(3)$ sigma model.  We may conjecture that at the centers of the spirals this value approaches a finite limiting value $f(0)\sim \pi/2$, i.e. for a critical solution the $O(3)$ field becomes restricted to a circle $S^1$. In the
spiral, the core of the configuration  
shrinks, however, the tail becomes more extended.  
At the same time, for the secondary branches of the spiral the numerical accuracy rapidly decreases, a different numerical approach is necessary for the study of this limit.

The contribution of the electromagnetic energy  modifies the $M(Q)$ dependency \cite{Kleihaus:2007vf}, see 
Fig.~\ref{fig2}, upper right plot. The spikes of the uncharged case transforms into a set of loops, the size of the loops increases with $e$. More importantly, depending on the
relative strength of gravity and the electromagnetic interaction, the gauged $O(3)$ boson stars exhibit different behavior  
in the limit
$\omega \to 1$.  

The gauged $O(3)$ configurations
experience
an additional repulsive interaction arising from the gauge
sector, as gravitational coupling  decreases, 
the relative contribution of the electrostatic energy becomes increasingly  essential. 

\begin{figure}[t!]
\begin{center}
\includegraphics[height=.35\textheight,  angle =-90]{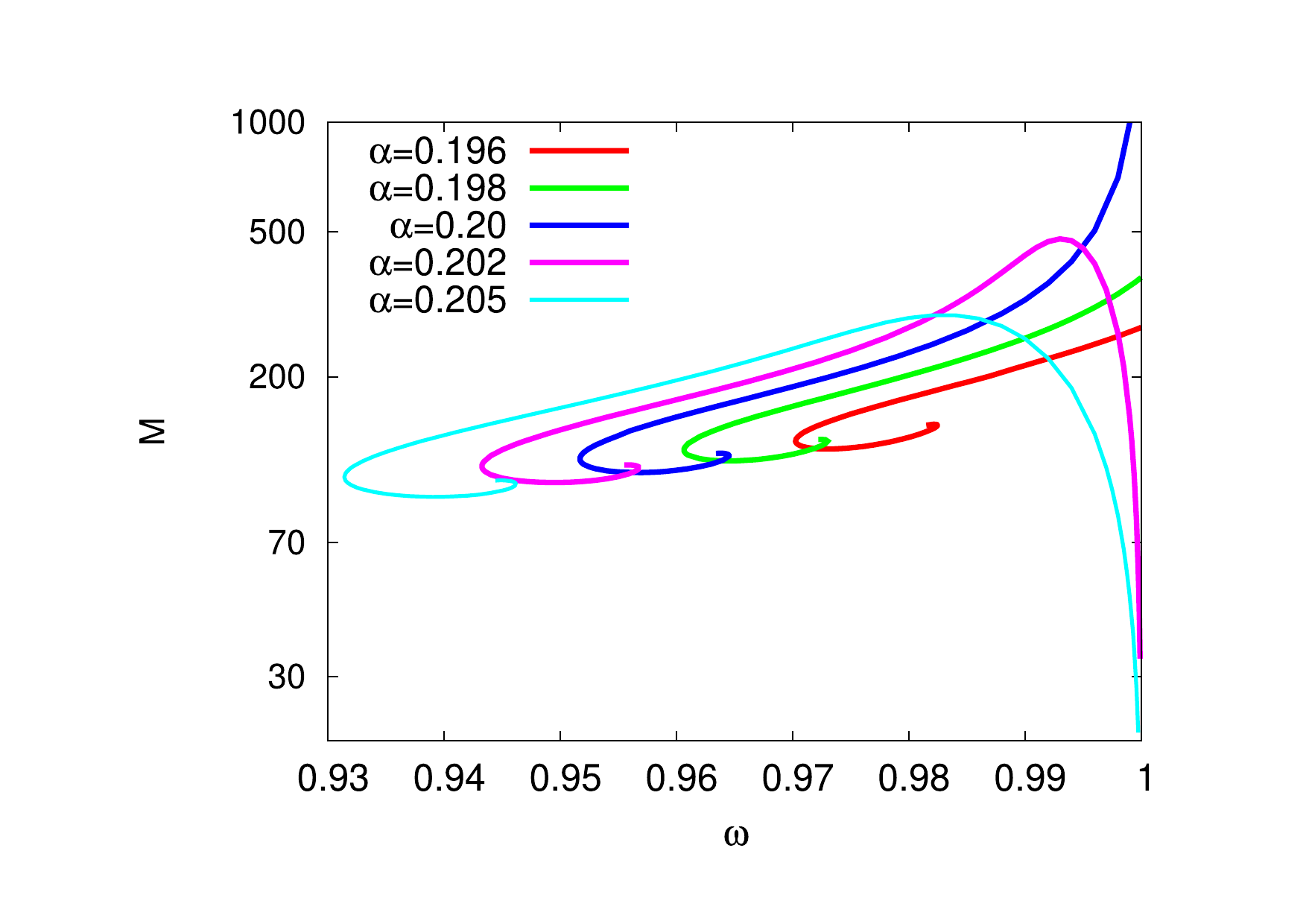}
\includegraphics[height=.35\textheight,  angle =-90]{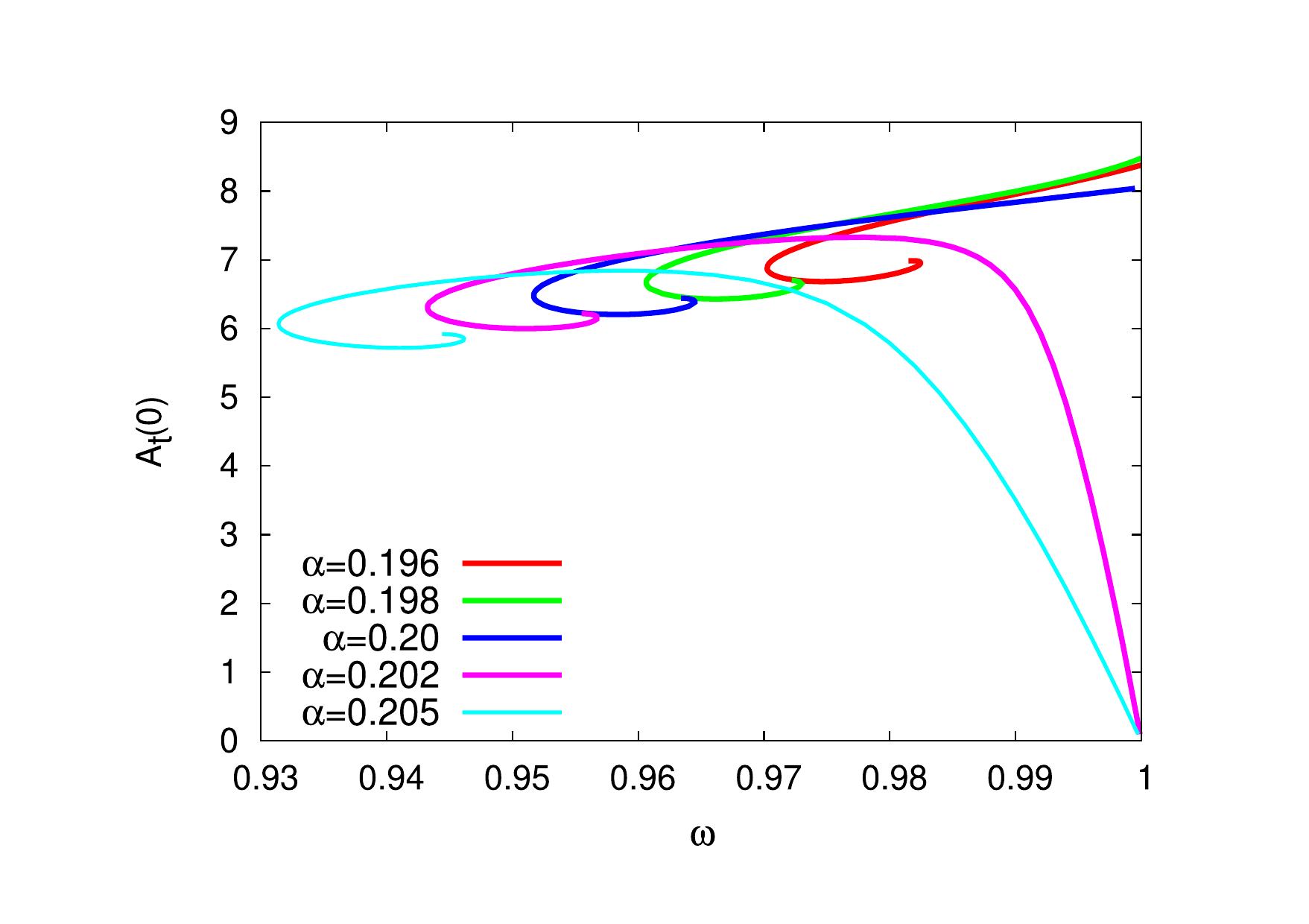}
\includegraphics[height=.35\textheight,  angle =-90]{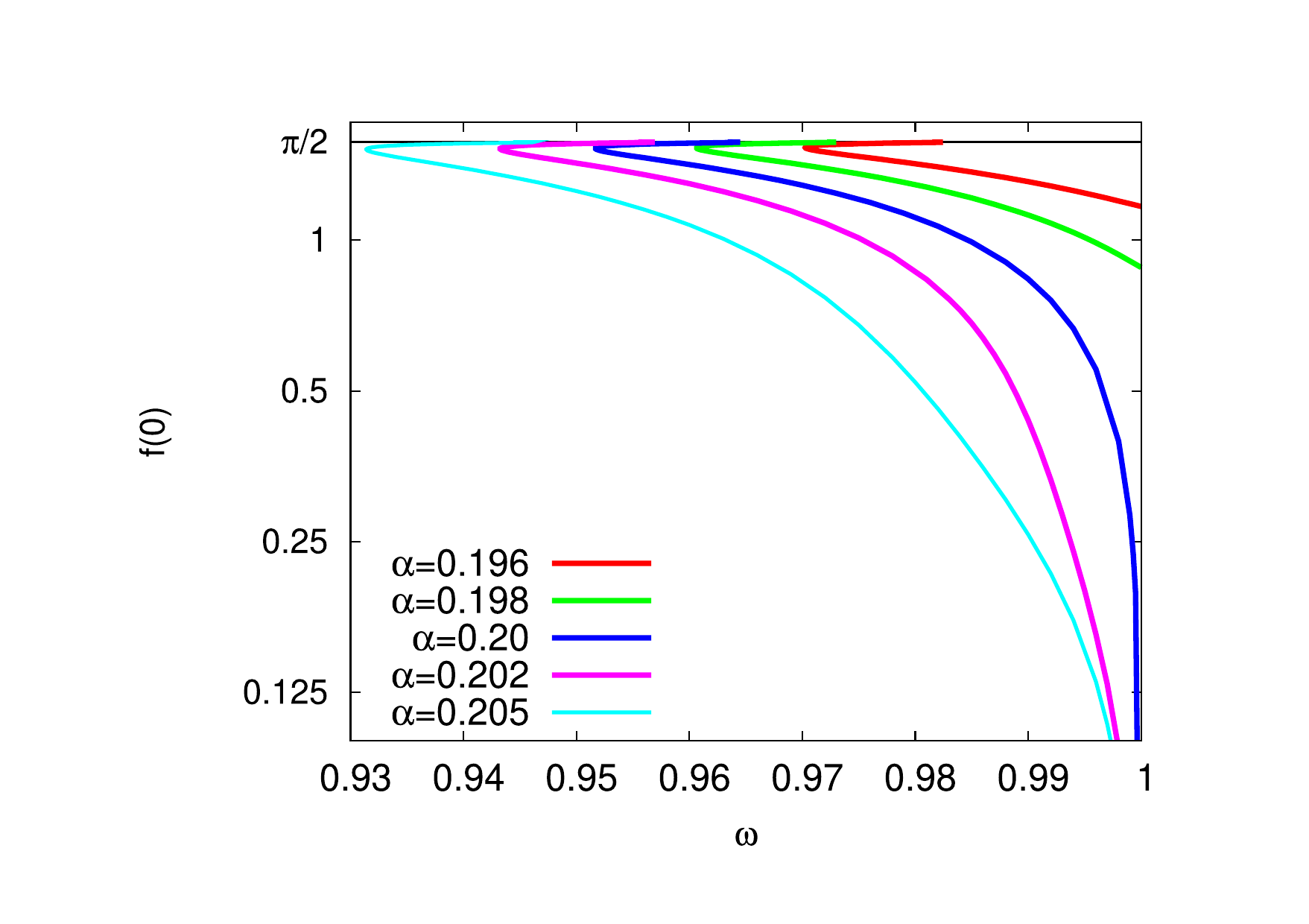}
\includegraphics[height=.35\textheight,  angle =-90]{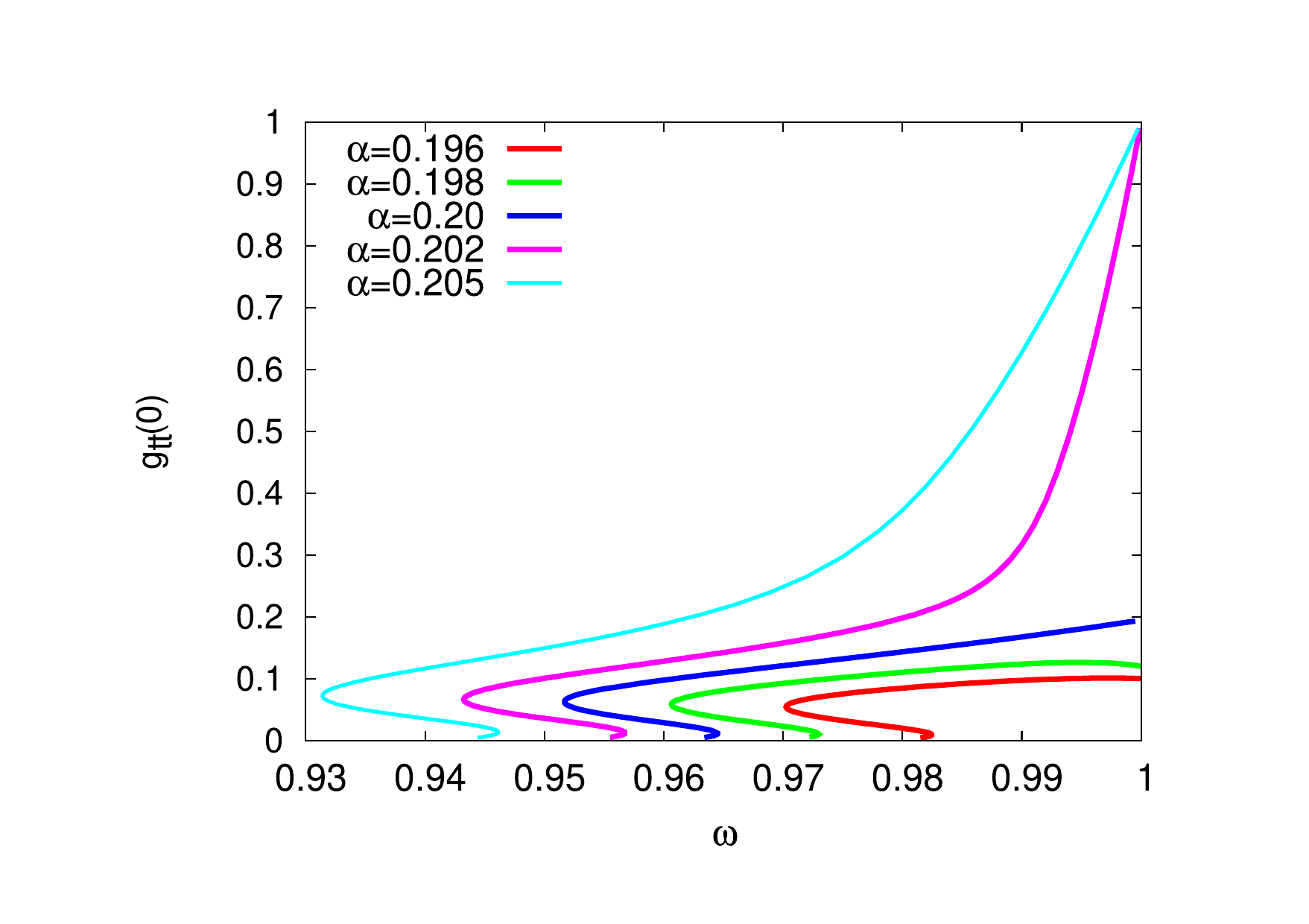}
\end{center}
\caption{\small Spherically symmetric $U(1)$ gauged gravitating $O(3)$ solitons: The ADM mass (upper left plot) in units of $8 \pi$, the value of the gauge potential $A_0$ at the origin  (upper right plot), 
the value of the scalar profile function $f$ at the origin (bottom left plot)
and the value of the metric function $g_{tt}$ at the origin  vs. frequency $\omega$ (bottom right plot)  are displayed for  some set of values of the gravitational coupling $\alpha$ and for the gauge 
coupling $e=0.1$.  
}    \lbfig{fig10}
\end{figure}

Given a value of the gravitational coupling $\alpha$, the minimal value of the angular
frequency $\omega_{min}$ increases with $e$, see Fig.~\ref{fig2}. For a given
value of the frequency $\omega$, charged soliton solutions appear to exist up to a maximal value of the gauge coupling constant only. Physically, this pattern is related with the electrostatic repulsion  which becomes stronger as $e$ increases. 

For a fixed value of the gauge coupling $e$, decrease of the gravitational coupling $\alpha$
entails an increase of the minimal
frequency 
$\omega_{min}$. 
Coincidentally,  the secondary branches of the spiral become shorter. Importantly, the limiting behavior of the fundamental branch at $\omega \to 1$ depends on the relative strength of the gauge and gravitational couplings. However, 
the  overall spiral  pattern is always observed. 
For sufficiently large gauge coupling  $e$ and sufficiently small gravitational coupling $\alpha$
the limit $\omega \to 1$ then is no longer Newtonian. This is illustrated in Fig.~\ref{fig10}. It appears that, for a given value of the gauge coupling, there is a critical value of the gravitational constant $\alpha_{cr}$, at which the repulsive force of electrostatic interaction is balanced by the gravitational attraction. In particular, for $e=0.1$ the critical gravitational coupling is $\alpha_{cr}\approx 0.20$. 

We notice that the usual Newtonian limit exists as $\alpha>\alpha_{cr}$. In this case, as $\omega \to 1$, the scalar field approaches the vacuum $\phi_\infty^a$, the metric attain flat space limit and the electrostatic potential is vanishing, see Fig.~\ref{fig10}. 
The critical case corresponds to 
divergent ADM mass $M$ and the charge $Q$,  
both the metric function $N$ and 
the electric potential $A_t$ approach some finite non-zero values it the center of configuration, although the scalar field trivializes and decouples from the gauge sector. 

A different scenario arises while $\alpha < \alpha_{cr}$, as demonstrated in Fig.~\ref{fig10}. Then the fundamental branch is dominated by the electromagnetic interaction, the limit $\omega  \to 1$  corresponds to the finite non-zero values of both the ADM mass $M$ and the charge $Q$, the scalar field is not in the vacuum,  the electric potential remains finite and the spacetime is  not flat. This pattern is qualitatively similar to that of the usual gauged boson stars \cite{Kunz:2021mbm}.

\subsection{Axially symmetric solutions (n=1)}
\begin{figure}[t!]
\begin{center}
\includegraphics[height=.28\textheight,  angle =-0]{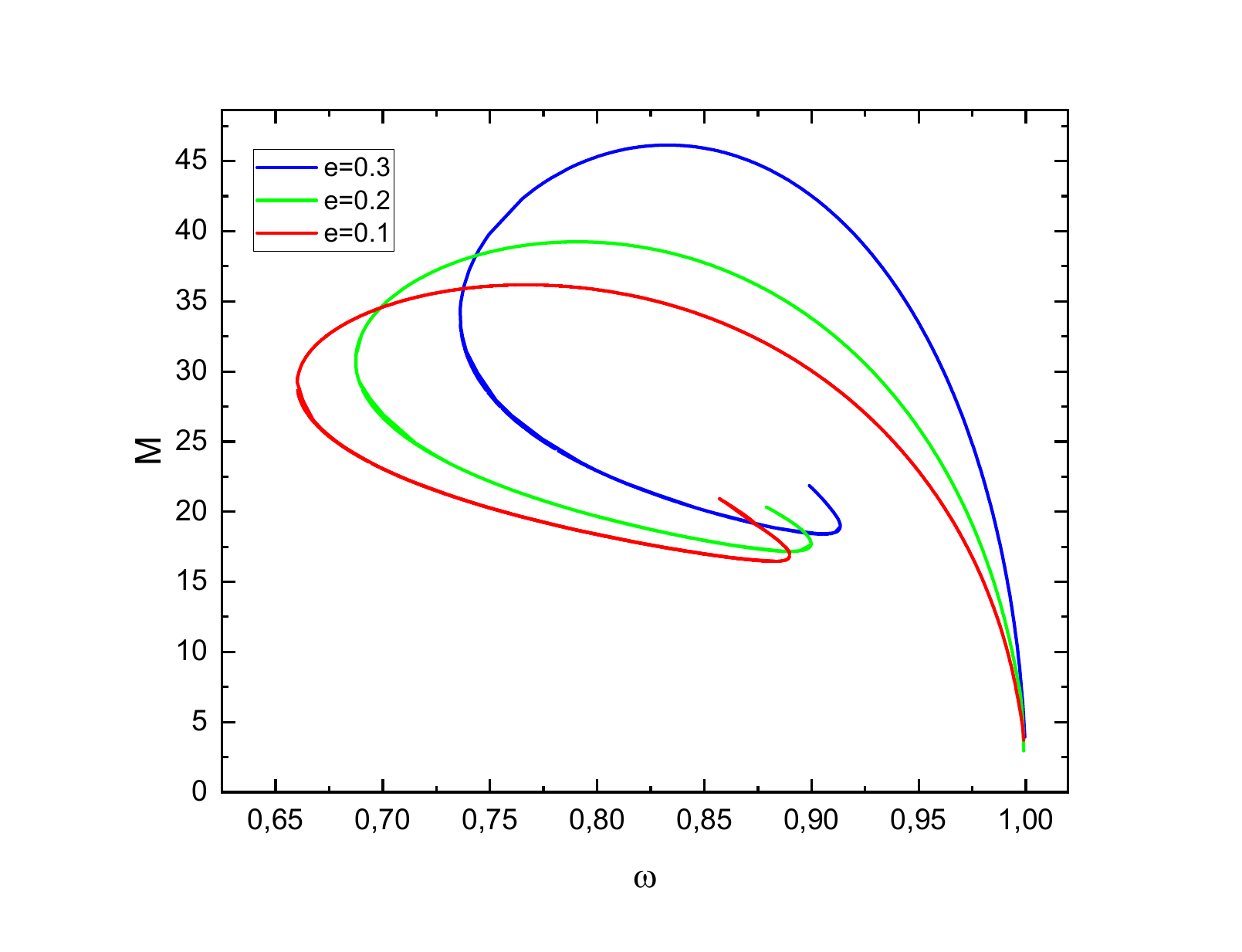}
\includegraphics[height=.28\textheight,  angle =-0]{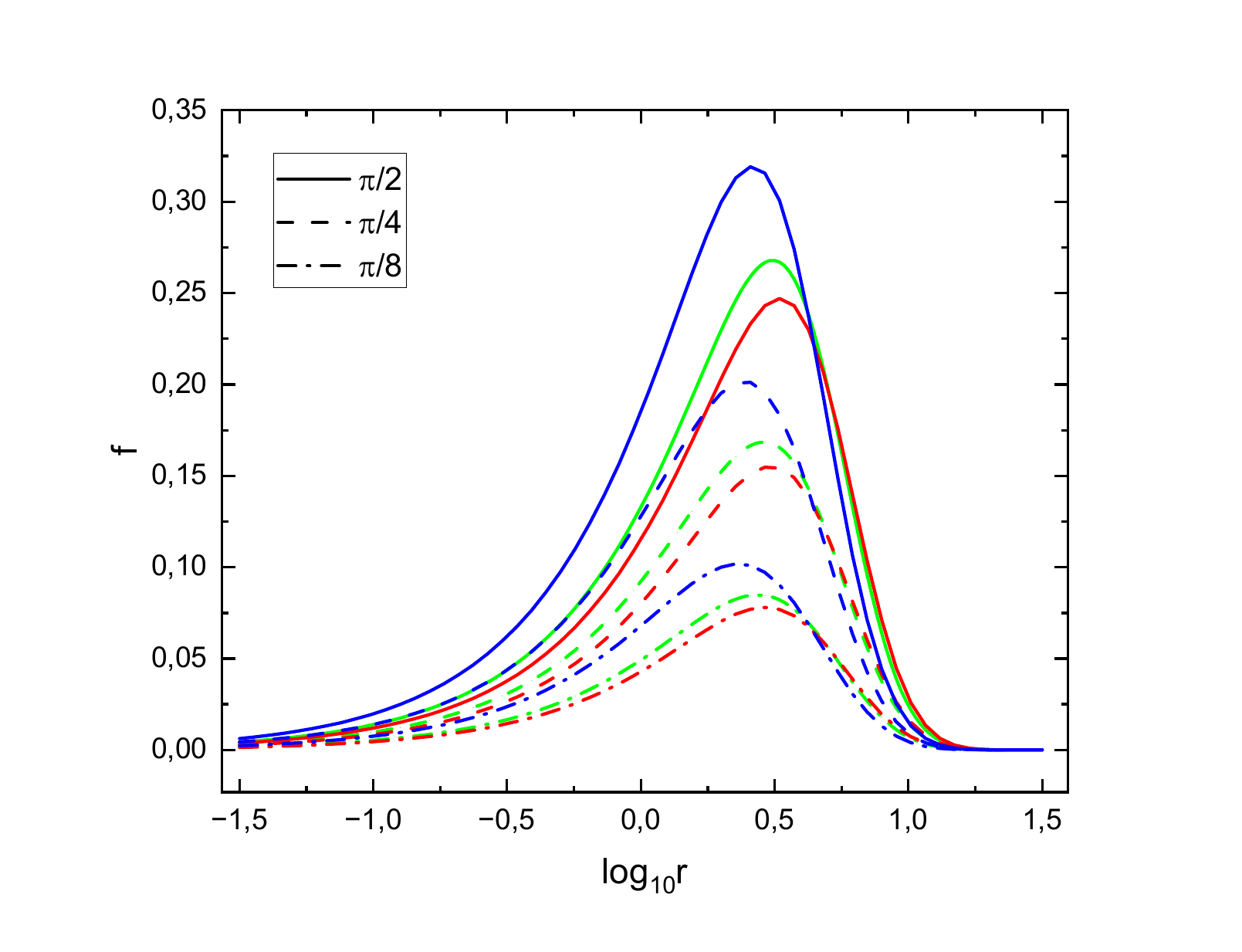}
\end{center}
\caption{\small Axially symmetric $U(1)$ gauged gravitating $O(3)$ solitons: The ADM mass (left) 
vs. frequency $\omega$ and  the 
scalar profile function $f$  
of illustrative solutions on the first branch at $\omega=0.85$
(right) are plotted for some set of values of the gauge coupling $e$ and for the gravitational coupling $\alpha^2=1$. 
}    \lbfig{fig3}
\end{figure}

\begin{figure}[h!]
\begin{center}
\includegraphics[height=.28\textheight,  angle =-0]{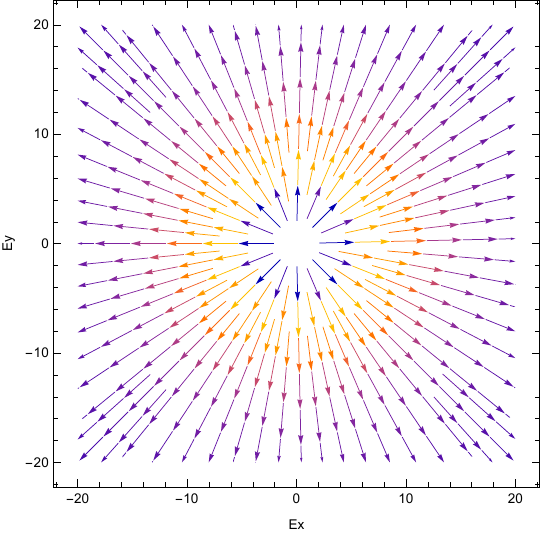}
\includegraphics[height=.28\textheight,  angle =-0]{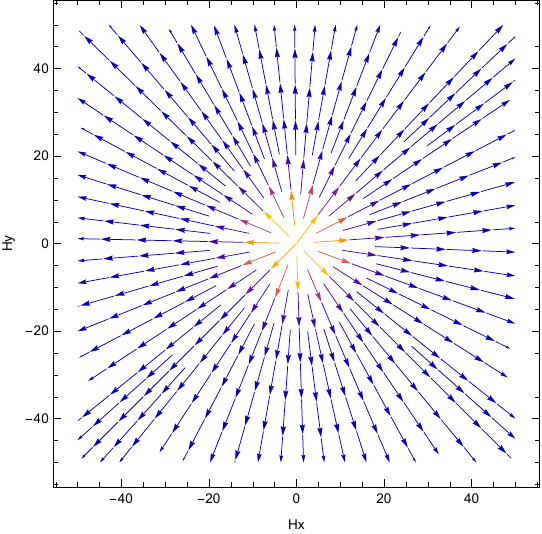}
\includegraphics[height=.28\textheight,  angle =-0]{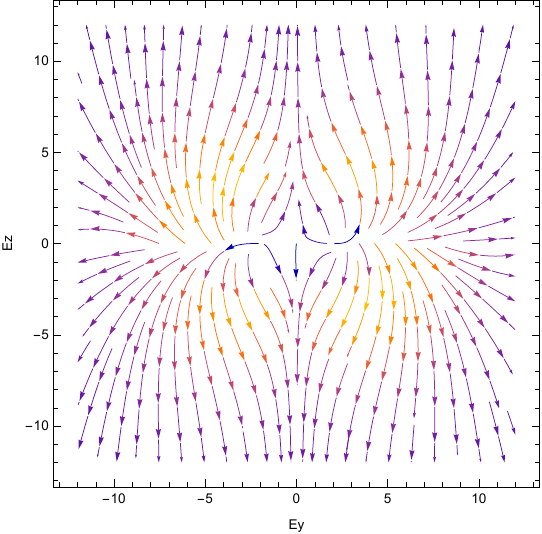}
\includegraphics[height=.28\textheight,  angle =-0]{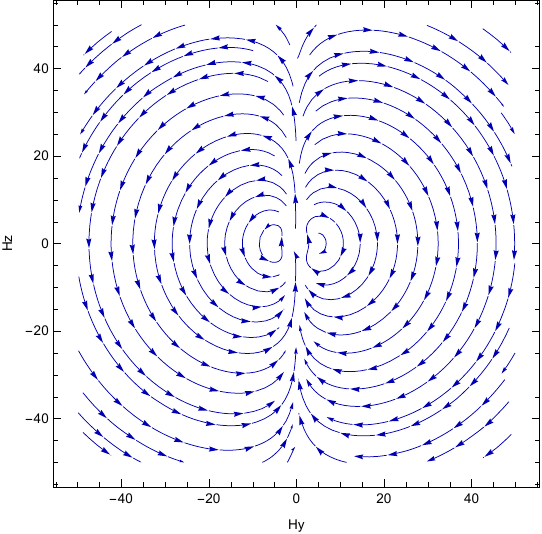}
\end{center}
\caption{\small $U(1)$ gauged gravitating $O(3)$ solitons: Electromagnetic field orientation of an illustrative axially-symmetric (n=1) solution on the first branch is displayed 
for gravitational coupling $\alpha^2=1$, frequency $\omega=0.85$, and gauge coupling $e=0.2$: the electric (upper left) and magnetic (upper right) field fluxes in the x-y plane, and 
the electric (bottom left) and magnetic (bottom right) field fluxes in the y-z plane.}    \lbfig{fig4}
\end{figure}

Spinning axially-symmetric $(n\ge 1)$ solitonic configurations are constructed as parity-even solutions of the system of equations \re{einsteq}, \re{eqfield} with the stationary ansatz \re{ansmetr},\re{anssfield} and \re{ansempot}, and the boundary conditions  \re{boundradinf},\re{boundang1},\re{boundang2} and \re{boundrad0}.  
We shall restrict our study to the case $n=1$, corresponding solutions are  qualitatively similar to those of higher values of the azimuthal winding. 

Apart from the sigma-model
constraint, charged axially symmetric gravitating solutions of the $O(3)$ model are akin to the
rotating $U(1)$ gauged boson stars  \cite{Collodel:2019ohy}. In both cases regular soliton solutions emerge similarly from the vacuum at the
maximal frequency $\omega_{max}=1$. 
The scalar profile function $f$ has a pronounced maximum on the symmetry plane, as displayed in  Fig.~\ref{fig3}, right plot. 
The amplitude of the field becomes larger as the gauge coupling becomes stronger (with the other parameters fixed).  

The dependency of these configurations on
the angular frequency $\omega$ is qualitatively similar to that just described for the spherically symmetric solitons, the ADM mass of the spinning $n=1$ solutions increases as the frequency decreases below the threshold, it approaches its maximum at
some value of the frequency $\omega< 1$, see  Fig.~\ref{fig3} left plot. The ADM mass and charge Q form a spiral, as $\omega$ is varied, while the metric function $f_0$ becomes exponentially small, see Fig.~\ref{fig5}, bottom plot. 
The maximum of
the scalar function $f$ on the equatorial plane
shows damped oscillations towards a maximal value, the amplitude of the axially-symmetric field is always smaller than the corresponding value of the 
fundamental spherically-symmetric solutions, see Fig.~\ref{fig5}, upper right plot. 

The total angular momentum of the axially-symmetric solutions of the $O(3) $ sigma model can be evaluated from \re{mass-mom}, as well as from the asymptotic decay of the metric function $W$. Recall that the charge $Q$ and the angular momentum are related via \re{Jn}.   

Axially symmetric configurations posses both electric and magnetic field, which is generated by the Noether current \re{ncurrent}. The corresponding toroidal magnetic field encircles the soliton, as seen in Fig.~\ref{fig4}.  The electric charge of the configuration is vanishing at the center of the spinning gauged $n=1$ soliton, the nodal structure of the electric field is resembling the field of a circular charged loop, the radius of the loop 
corresponds to the characteristic size of the configuration, for a given value of the gauge coupling, it increases on the first fundamental branch of solutions until the maximal mass is attained; then the trend inverts and the size of the soliton decrease until the minimum ADM
mass attained along the second branch. The
solutions then follows a spiraling/oscillating
pattern, with successive backbendings towards 
a limiting singular solution at the center of the spiral, which, however, possesses finite mass and charge. 
As before, the maximal value of the scalar field is restricted by the sigma-model constrain while it diverges for the usual boson stars.  

\begin{figure}[t!]
\begin{center}
\includegraphics[height=.28\textheight,  angle =-0]{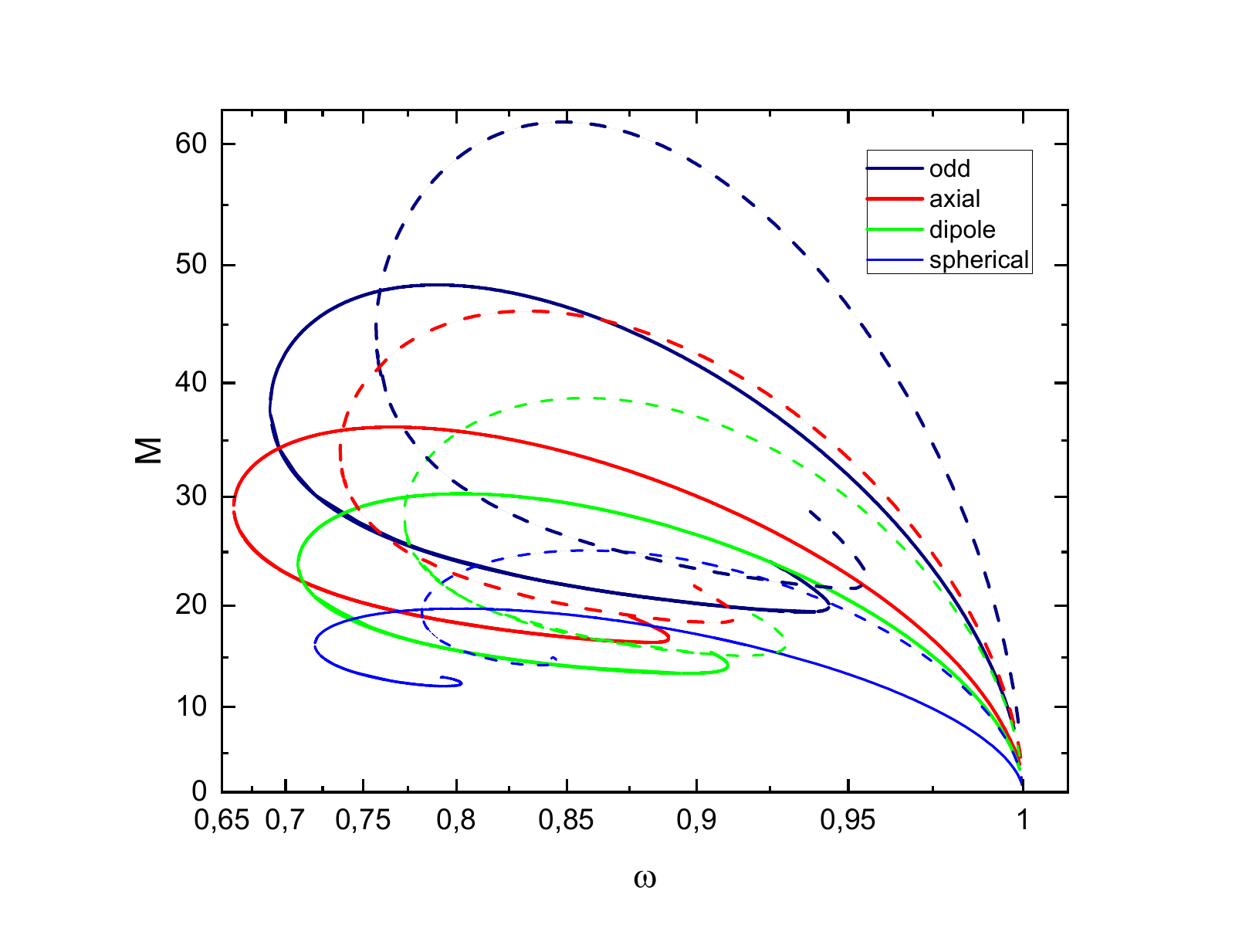}
\includegraphics[height=.28\textheight,  angle =-0]{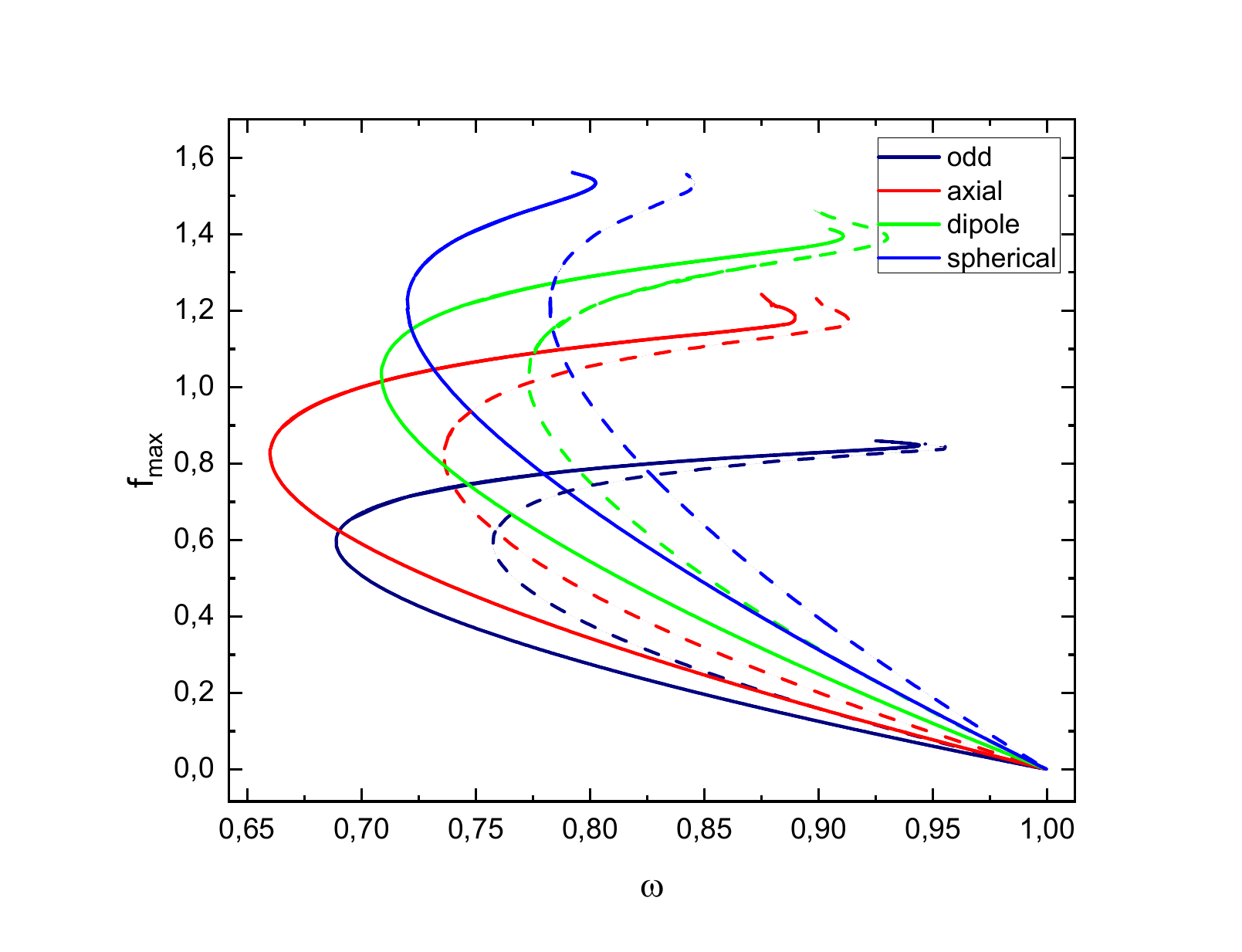}
\includegraphics[height=.28\textheight,  angle =-0]{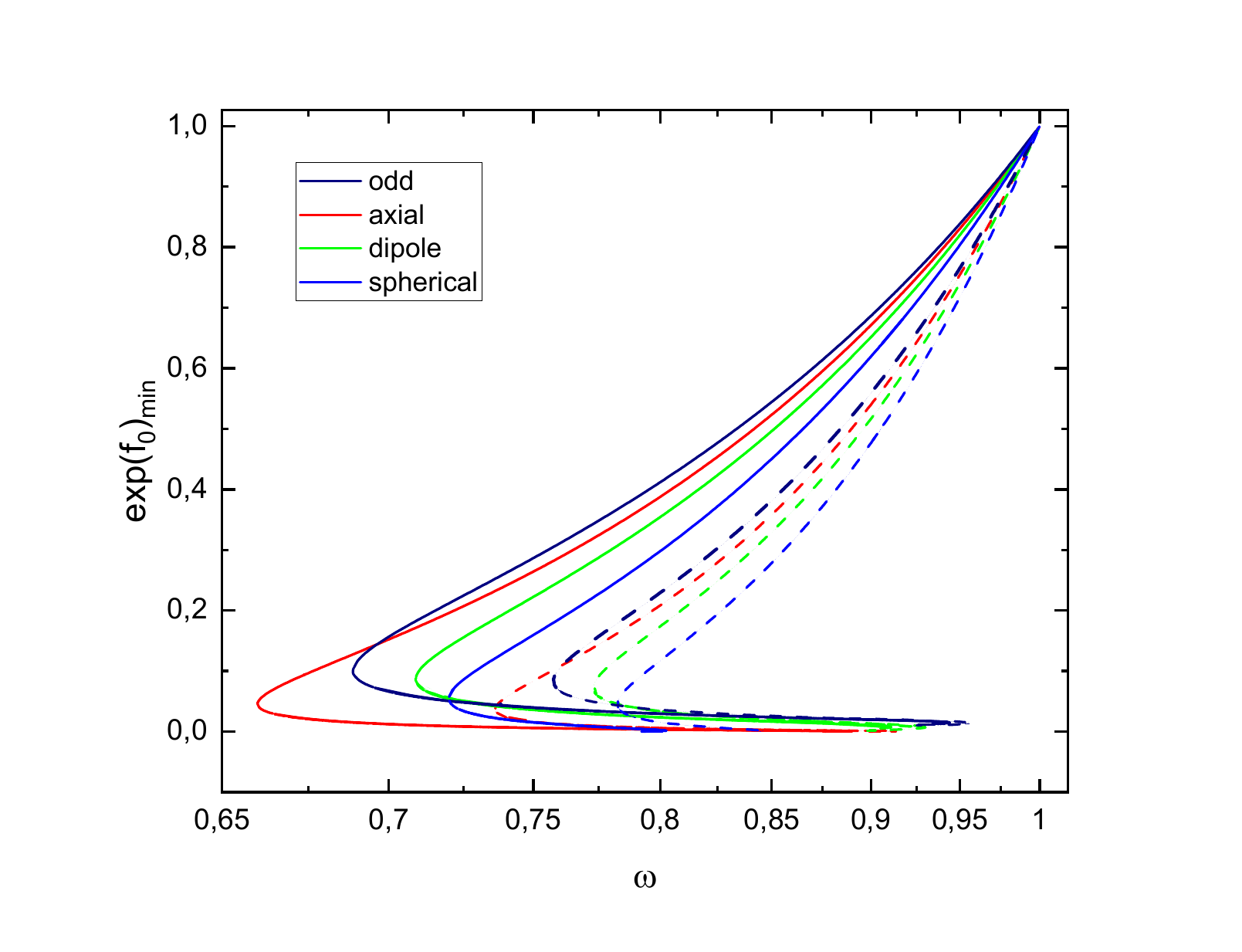}
\end{center}
\caption{\small Different types of $U(1)$ gauged gravitating $O(3)$ solitons: The ADM mass (upper left), the maximal value of the 
scalar profile function $f$ (upper right) and the exponent of the metric function $f_0$ (bottom plot) are plotted vs. the frequency $\omega$ for two values of the gauge coupling $e=0.1$ (solid line) and $e=0.3$ (dashed line) and for gravitational coupling $\alpha^2=1$. 
}    \lbfig{fig5}
\end{figure}

\subsection{Parity-odd solutions (n=1)}
By analogy with axially symmetric boson stars and spinning Q-balls in flat space \cite{Kleihaus:2005me,Kleihaus:2007vk,Radu:2008pp,Kunz:2019bhm}, 
for each non-zero value of the integer winding number $n$, there should be two types of spinning solutions of the non-linear $O(3)$ sigma-model possessing different parity, the parity-even and the parity-odd, respectively. Similar to the solutions of other types, the parity-odd configurations do not possess the flat space limit.

For the parity-odd solutions the scalar $O(3)$ field vanishes both on the z-axis and
in the xy-plane, the energy density distribution exhibits a double torus-like
structure, with two tori located symmetrically with respect to the equatorial plane, as illustrated in Fig.~\ref{fig1}.

The dependence of these solutions on the frequency $\omega$ is
qualitatively similar to that of the other spinning boson stars, as seen in Fig.~\ref{fig5}. As expected, for
the same set of values of the parameters of the model, the mass of the parity-odd configurations is higher than the
mass of corresponding solutions of other types, see Fig.~\ref{fig5}, upper left plot. The minimal frequency $\omega_{min}$ of the parity-odd configurations is higher than the analogous minimal frequency of the fundamental spherically-symmetric solution but smaller than the minimal frequency of the parity-even $n=1$ axially-symmetric configurations. 

The  mass of the parity-odd solution also exhibit a spiral structure towards a limiting solution.  The spatial distribution of the $O(3)$ scalar field on the first branch exhibit two local extrema at $\theta = \pi/4,~ 3\pi/4$, see Fig.~\ref{fig7}, left plot . As usual, the emergency of the frequency-mass spiral may be related to damped oscillations in the force-balance between the repulsive scalar interaction and the gravitational attraction in equilibrium. The electromagnetic interaction shifts the balance. In the spiral the gravitational attraction becomes stronger, the characteristic size of the localized configuration decreases. Interestingly, the shape of the distribution of the scalar field changes along the third branch of the spiral, it becomes more compact and a quadruple torus-like structure arises on the forth branch, see Fig.~\ref{fig7}. 
This observation suggests that multi-torus configurations may arise on subsequent branches of the spiral. 

\begin{figure}[thb]
\begin{center}
\includegraphics[height=.36\textheight,  angle =-90]{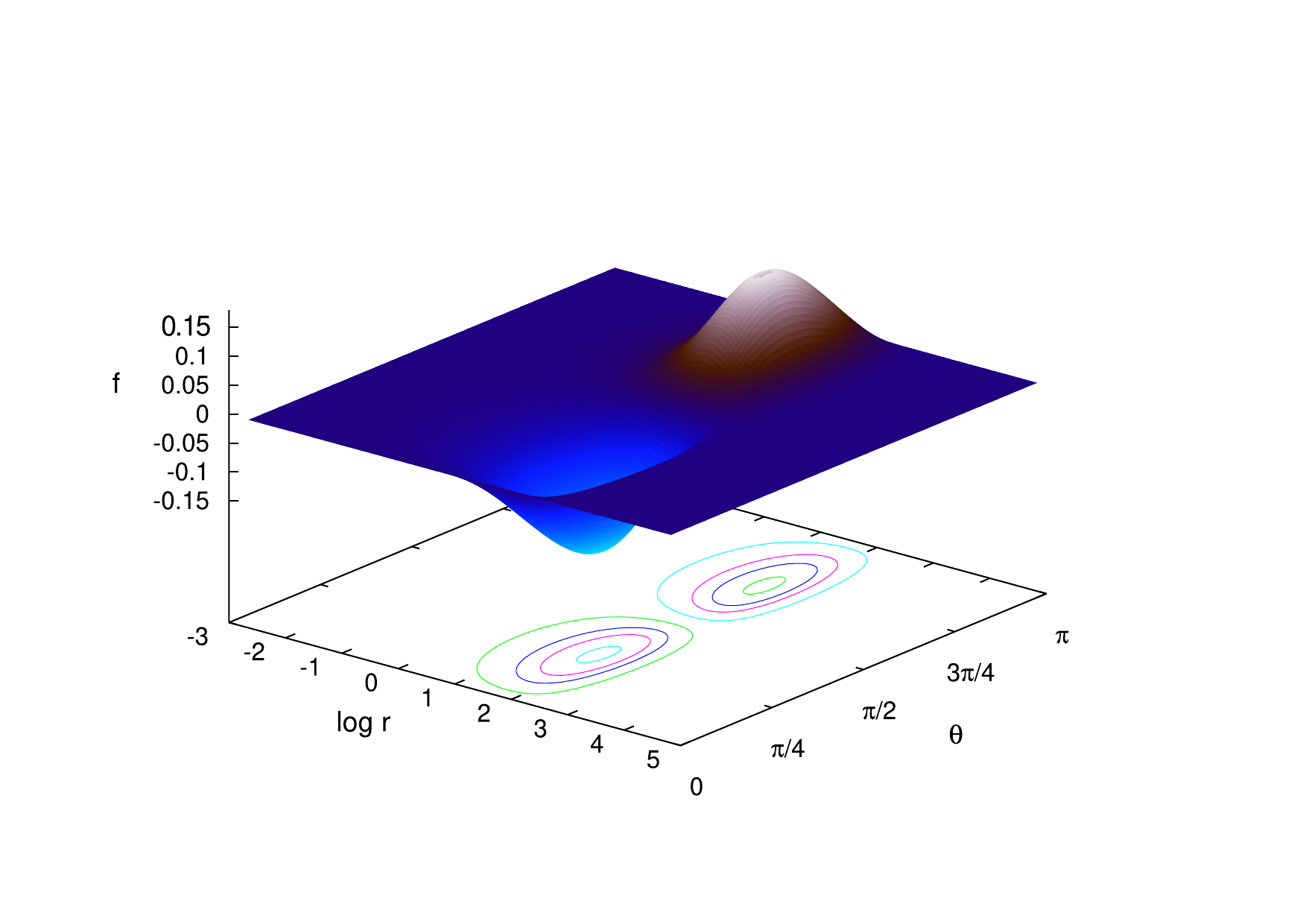}
\includegraphics[height=.36\textheight,  angle =-90]{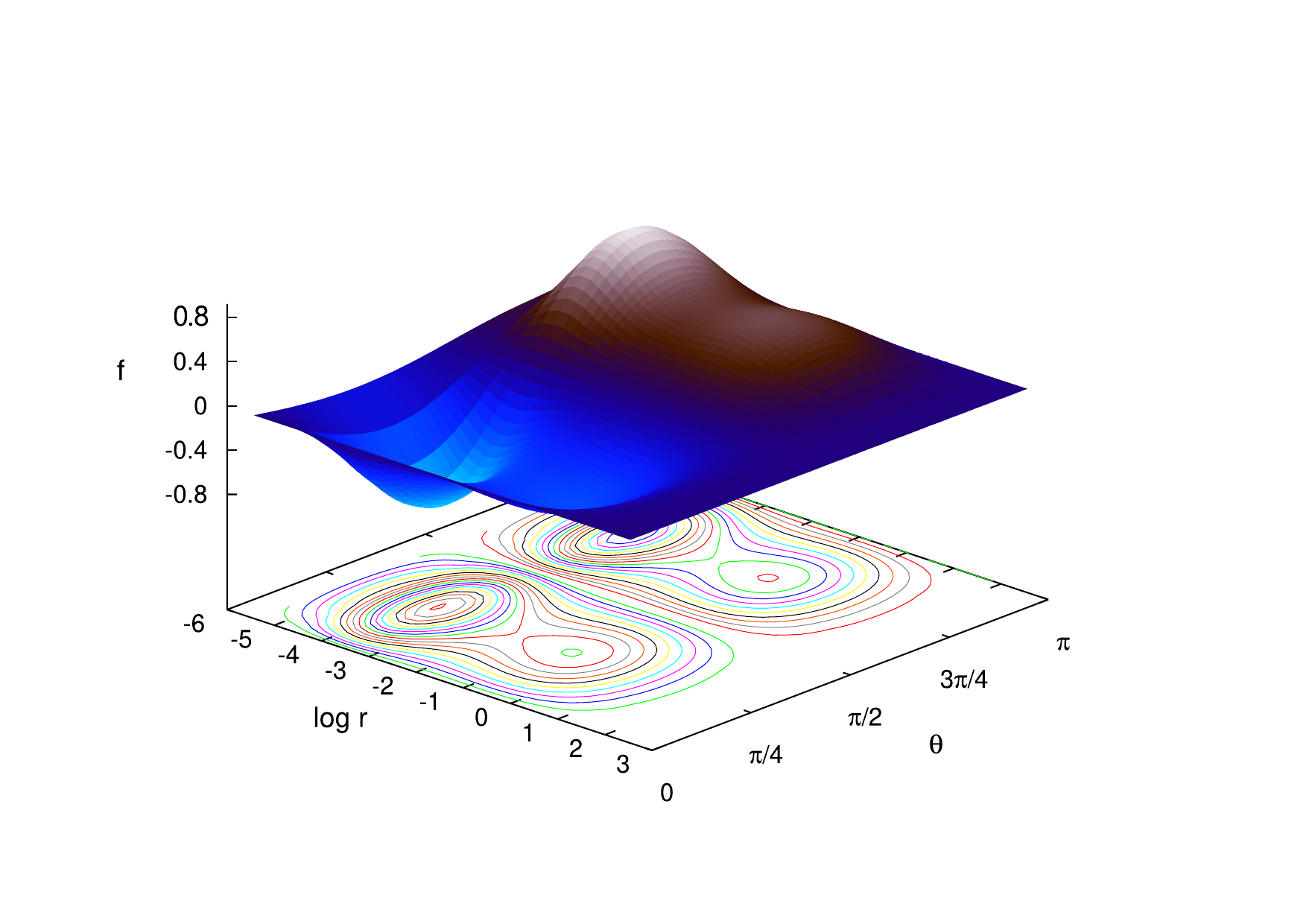}
\end{center}
\caption{\small The $O(3)$ scalar field for
$n=1$ negative parity solutions for 
the gauge coupling $e=0.1$ and  gravitational coupling $\alpha^2=1$ at $\omega=0.86$ on the first (left) and the forth (right) branches versus the
coordinates $\log r$ and $\theta$. 
}    \lbfig{fig7}
\end{figure}

The maximal value of the scalar amplitude always remains smaller than for solutions of other types, approximately half the respective maximal value of the fundamental spherically-symmetric solution, see Fig.~\ref{fig5}, upper right plot. Again, increase of the gauge coupling is related with a growing
contribution of the electromagnetic forces, 
the secondary branches become less extended.

\begin{figure}[h!]
\begin{center}
\includegraphics[height=.28\textheight,  angle =-0]{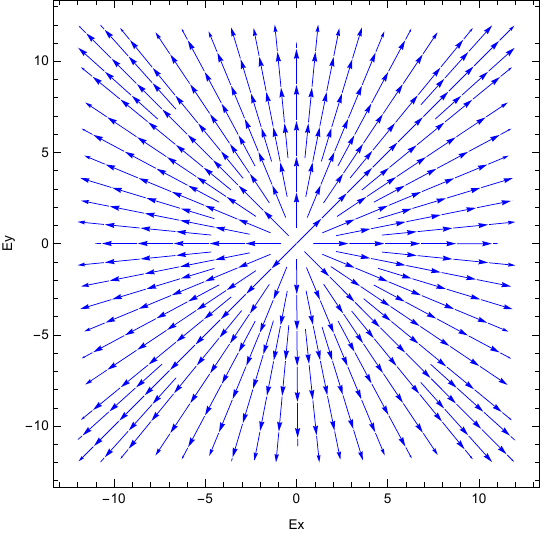}
\includegraphics[height=.28\textheight,  angle =-0]{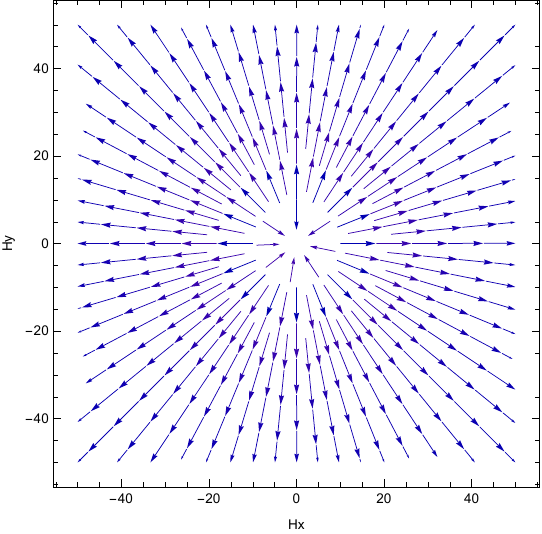}
\includegraphics[height=.28\textheight,  angle =-0]{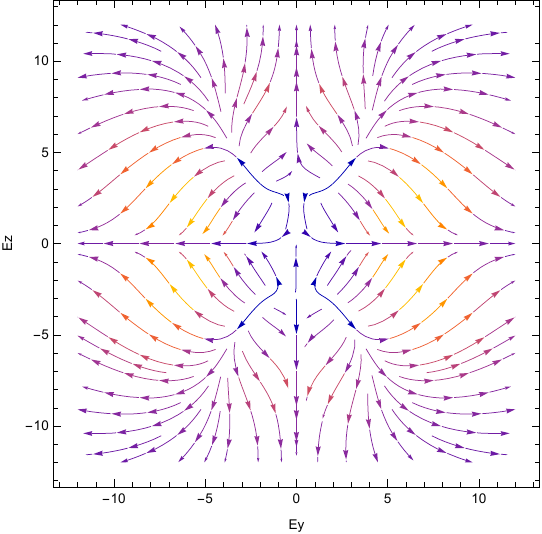}
\includegraphics[height=.28\textheight,  angle =-0]{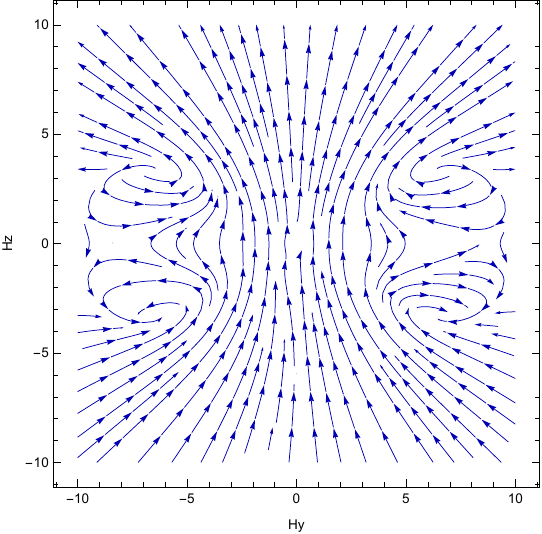}
\end{center}
\caption{\small $U(1)$ gauged gravitating $O(3)$ solitons: Electromagnetic field orientation of an illustrative parity-odd (n=1) solution on the first branch is displayed 
for gravitational coupling $\alpha^2=1$, frequency $\omega=0.85$, and gauge coupling $e=0.2$: the electric (upper left) and magnetic (upper right) field fluxes in the x-y plane, and 
the electric (bottom left) and magnetic (bottom right) field fluxes in the y-z plane.}    \lbfig{fig6}
\end{figure}

The $U(1)$ gauged parity-odd solutions posses both electric and magnetic field, which is generated by the Noether current \re{ncurrent}.  
The configuration is encircled by a double toroidal flux, as 
displayed in Fig.~\ref{fig6}, right plots.
The corresponding electric field exhibit peculiar structure, see Fig.~\ref{fig6}, left plots. It can be interpreted as being generated by two charged discs placed symmetrically with respect to the equatorial plane.   

\subsection{Dipole solutions (n=0)}
We now turn to the dipole solution, which represents two non-spinning, $U(1)$ gauged $O(3)$ boson stars in equilibrium. These configurations resemble the corresponding two-center dipolar boson stars \cite{Yoshida:1997nd,Herdeiro:2021mol,Cunha:2022tvk,Ildefonso:2023qty}, they exist due to the balance between the gravitational attraction and the (short-range)
scalar mediated repulsion. Long-range electrostatic interaction between the gauged $O(3)$ solitons shifts this balance, each component of the dipole possesses an electric charge, as
shown in Fig.~\ref{fig9}. 
For a given value of the gravitational coupling $\alpha$ 
an increase of the
gauge coupling $g$ entails an increase of the electric charge and, subsequently, an increase 
of the equilibrium distance between the components. 

\begin{figure}[t!]
\begin{center}
\includegraphics[height=.35\textheight,  angle =-90]{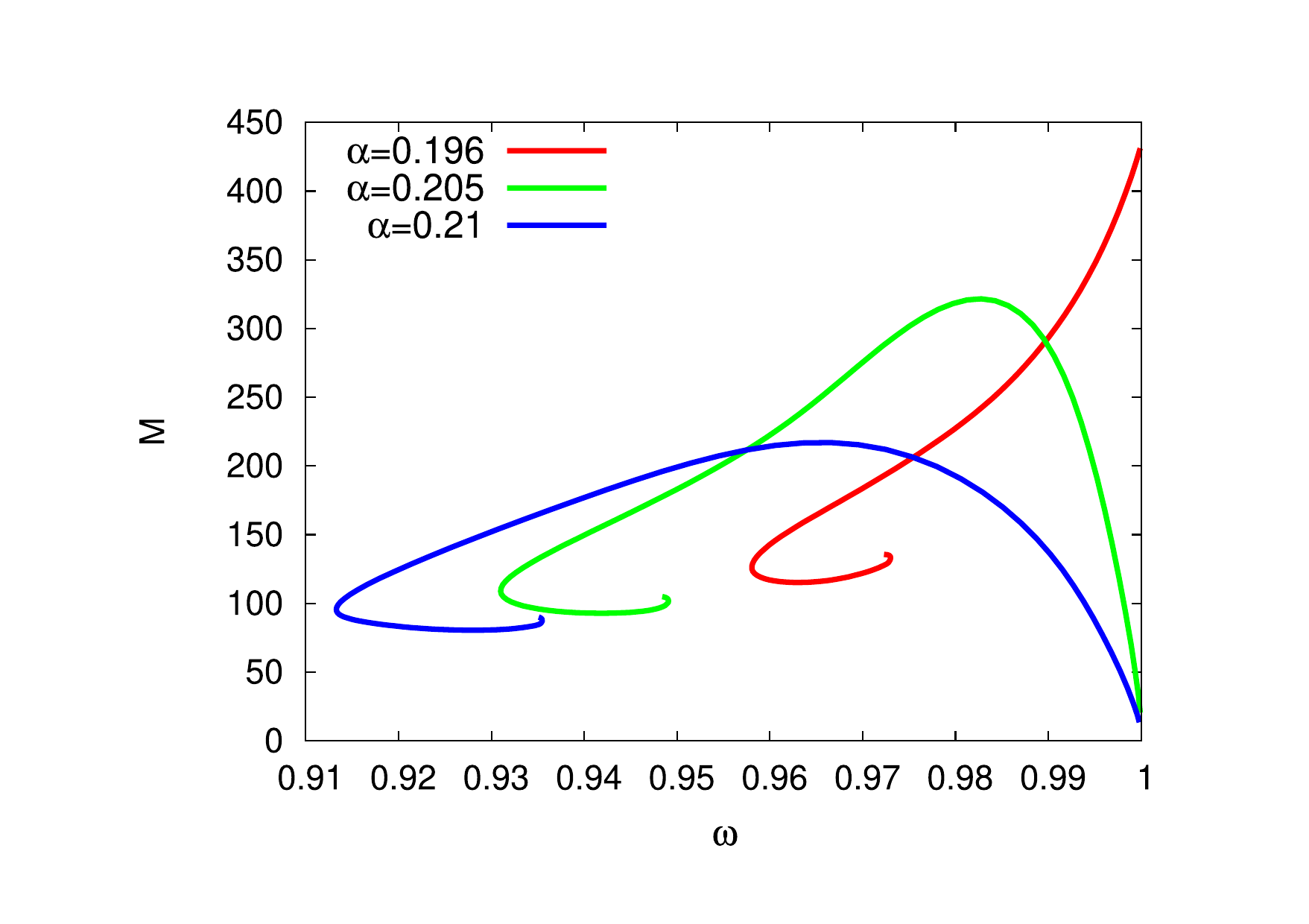}
\includegraphics[height=.35\textheight,  angle =-90]{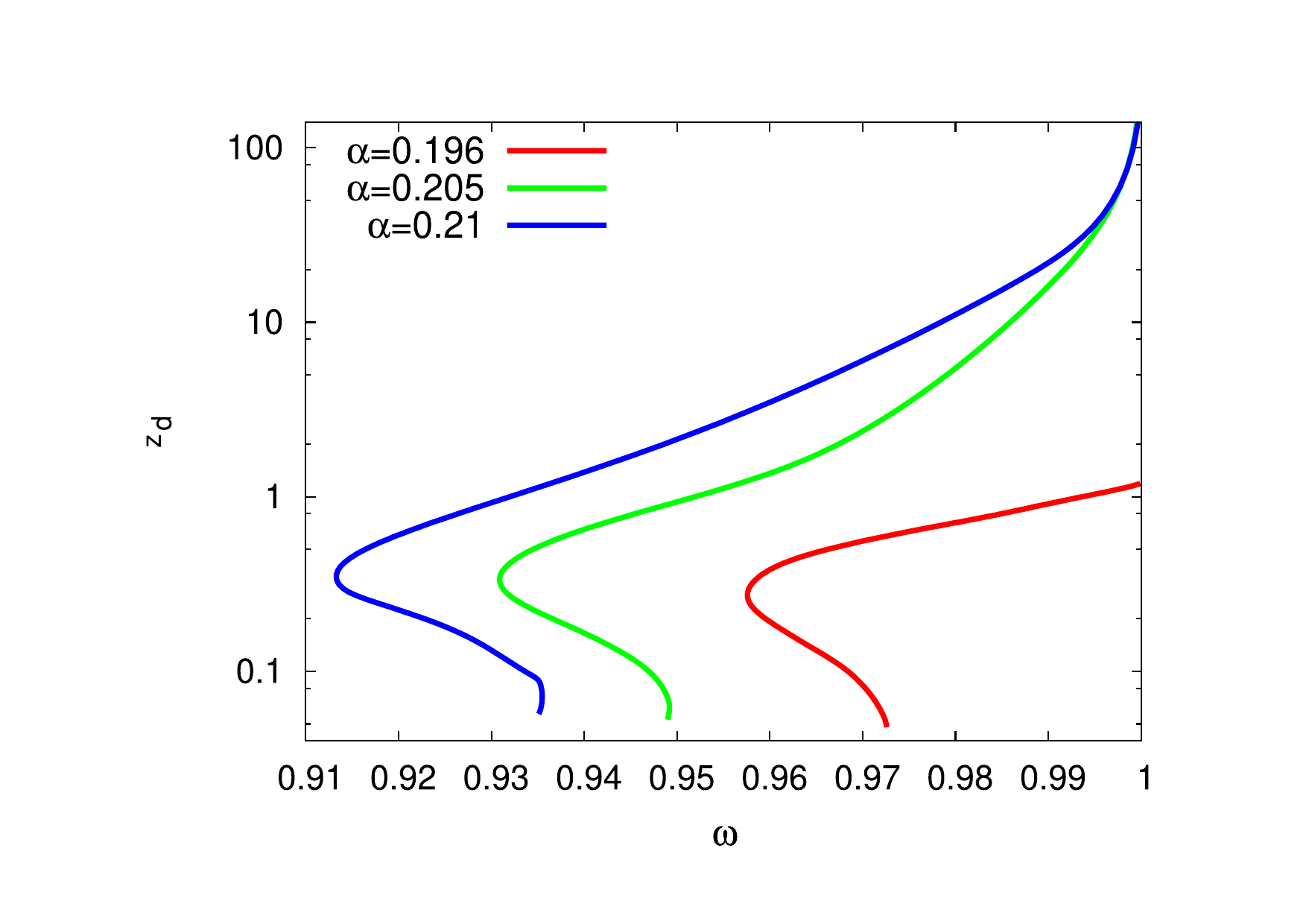}
\includegraphics[height=.35\textheight,  angle =-90]{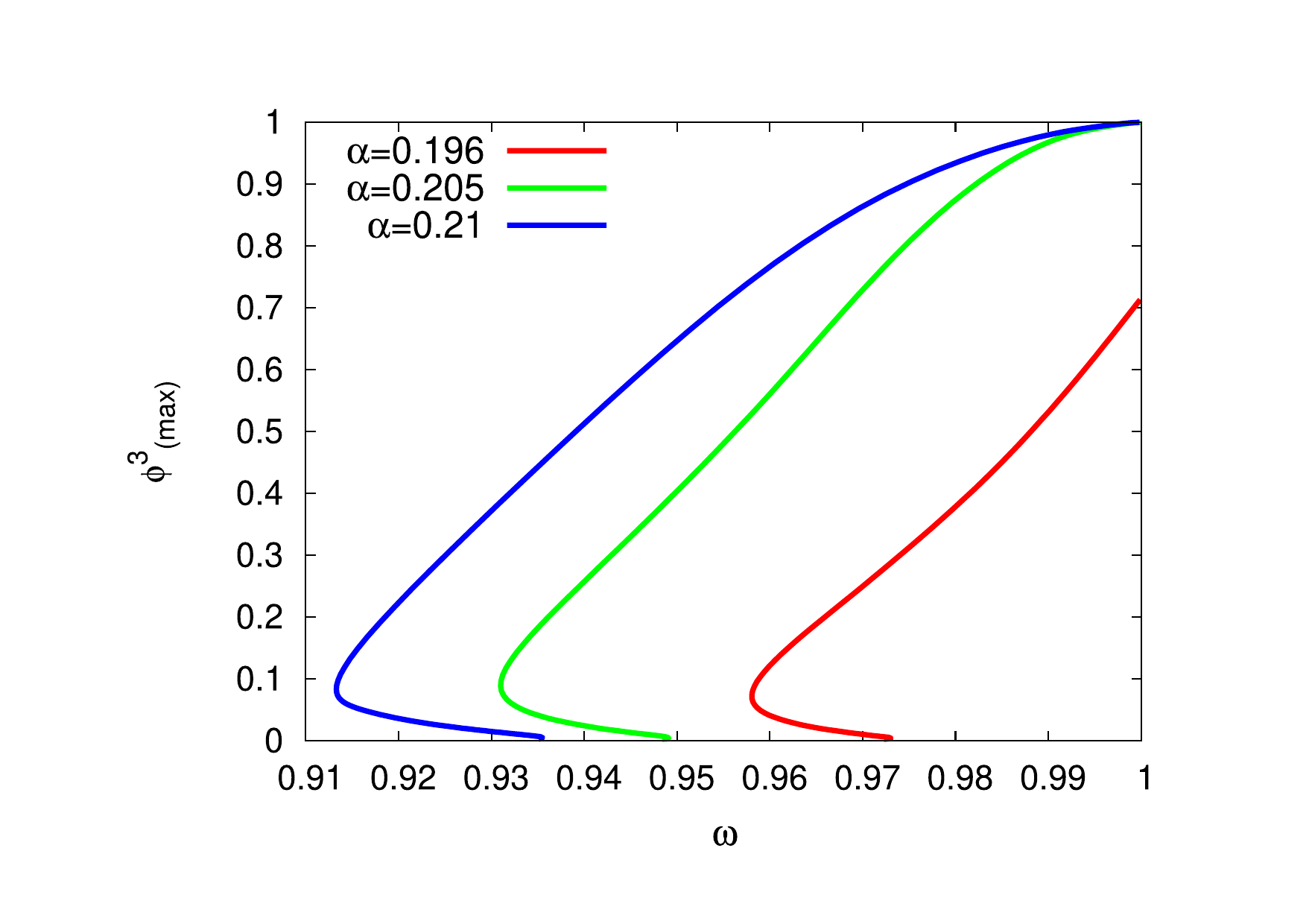}
\includegraphics[height=.35\textheight,  angle =-90]{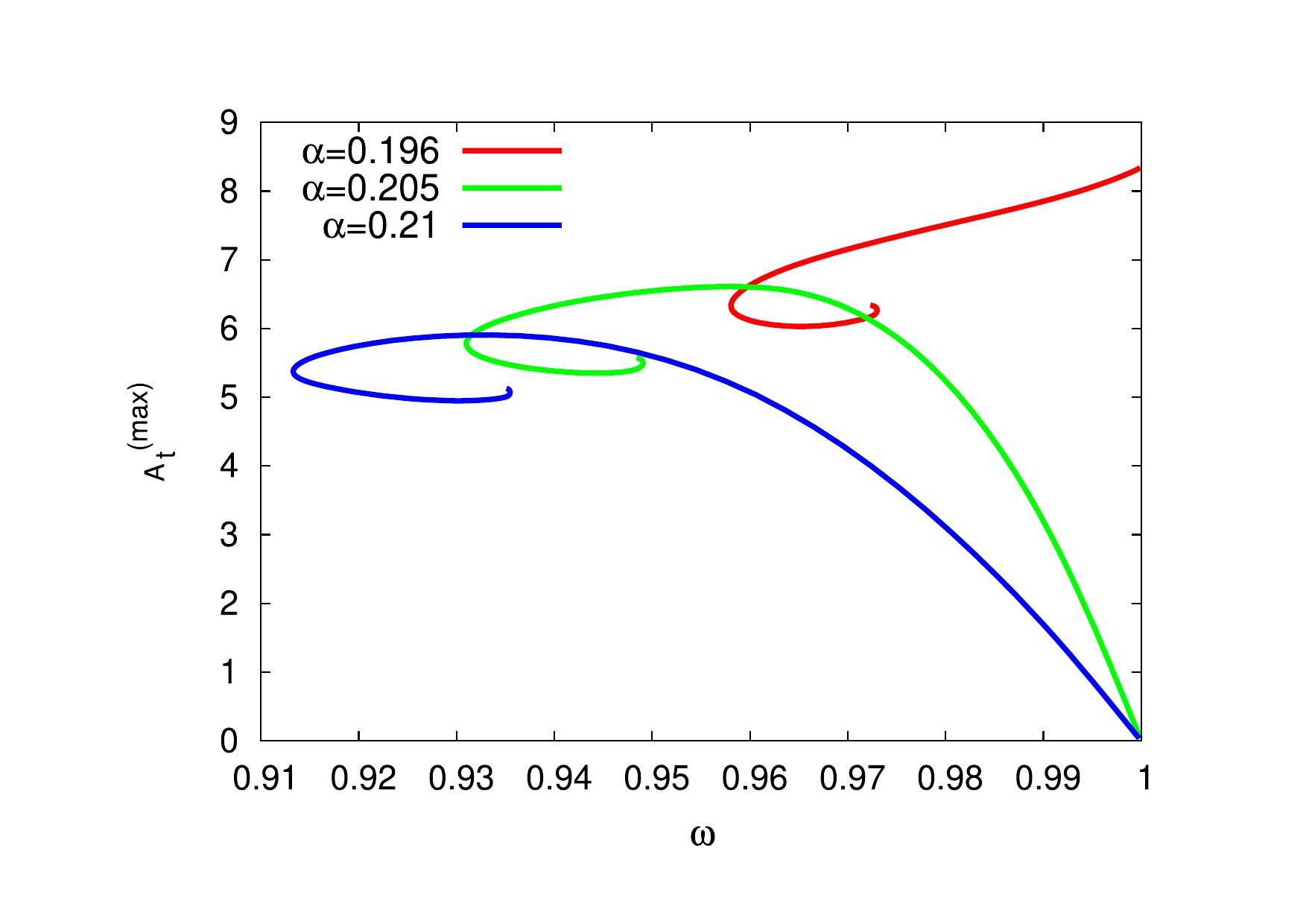}
\end{center}
\caption{\small  $U(1)$ gauged gravitating $O(3)$ dipole solitons: The ADM mass (upper left plot) in units of $8 \pi$, the separation $z_d$ between the two components (upper right plot), 
the maximal value of the scalar component $\phi^3$ (bottom left plot) and the maximal value of the 
gauge potential $A_0$  (bottom right plot)
vs. frequency $\omega$ are displayed for several values of the gravitational coupling $\alpha$ for the gauge 
coupling $e=0.1$.  
}    \lbfig{fig11}
\end{figure}

The dipole solution can be considered as a limiting non-rotating ($n=0$) parity odd configuration of self-gravitating scalar field. Two
components of the dipole are individual electrically charged boson stars with a phase difference of $\pi$ \cite{Herdeiro:2021mol,Cunha:2022tvk}. 
Indeed, inversion of the sign of the scalar field function under
reflections $\theta \to \pi - \theta$  corresponds to the shift of the
phase $\omega t \to \omega t - \pi$.
Also, it was pointed out that the character of the interaction
between Q-balls in Minkowski spacetime depends on their relative phase, it is attractive is the Q-balls are in phase, and it is repulsive if the Q-balls are in opposite phases \cite{Battye:2000qj}. 

\begin{figure}[thb]
\begin{center}
\includegraphics[height=.28\textheight,  angle =-0]{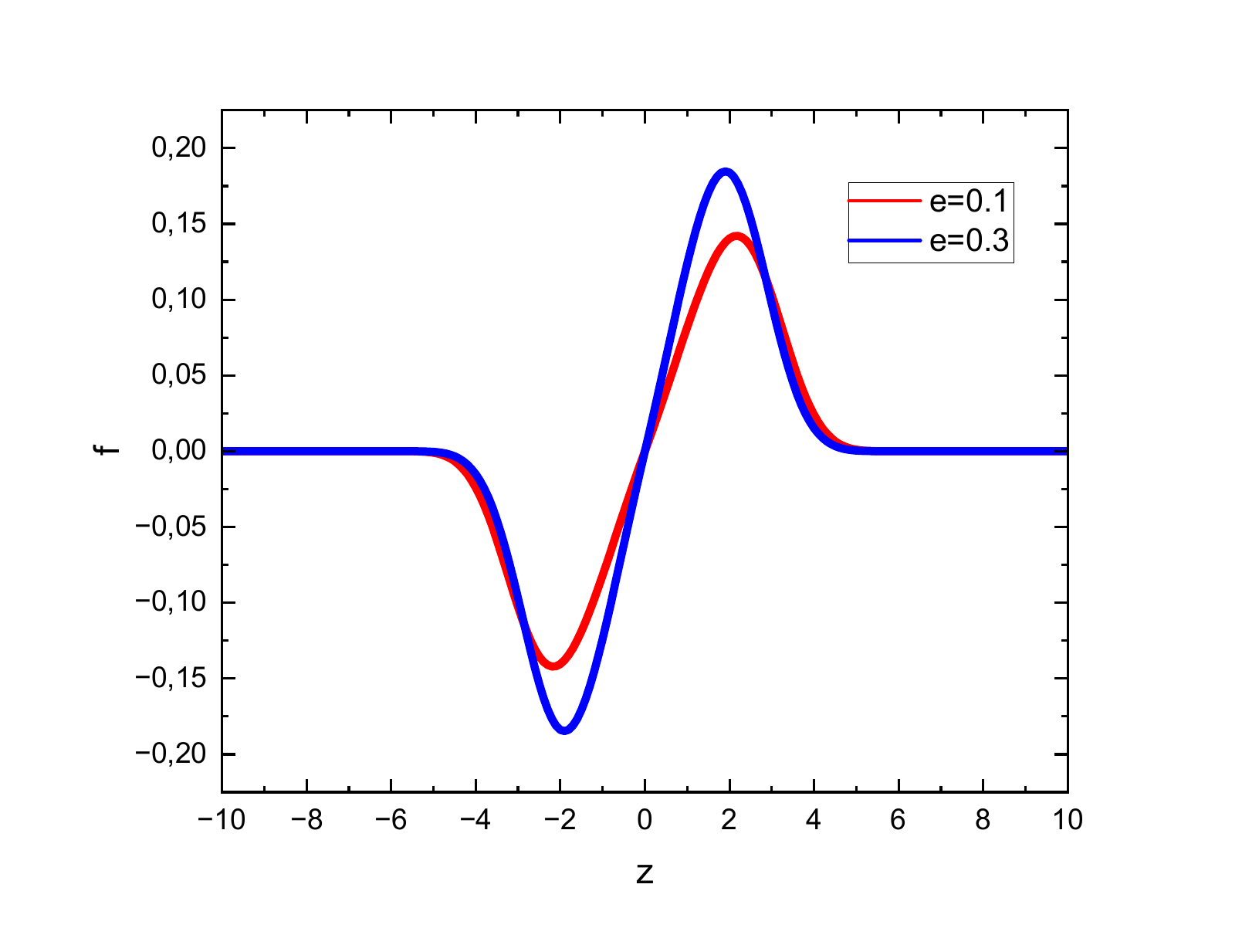}
\includegraphics[height=.28\textheight,  angle =-0]{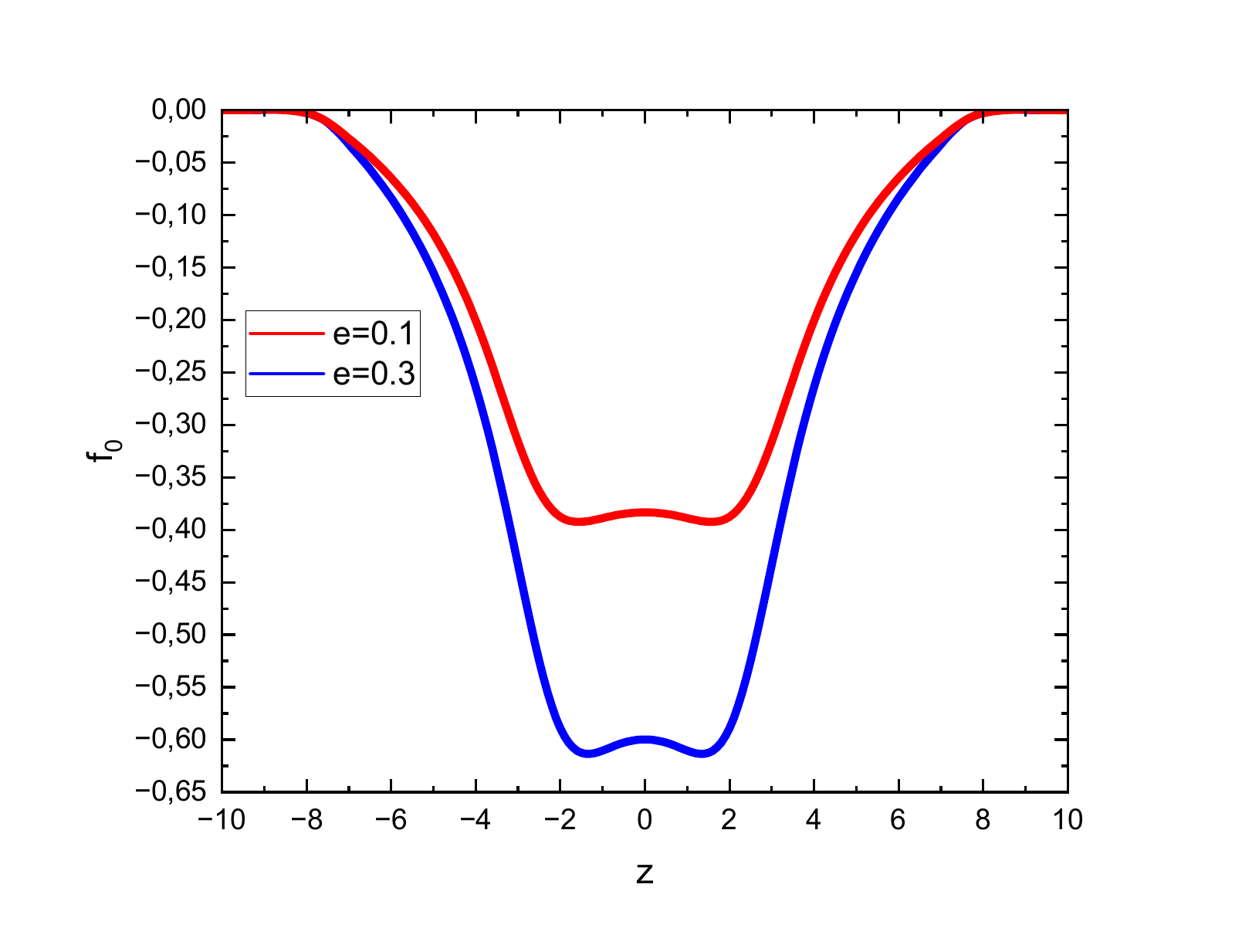}
\includegraphics[height=.28\textheight,  angle =-0]{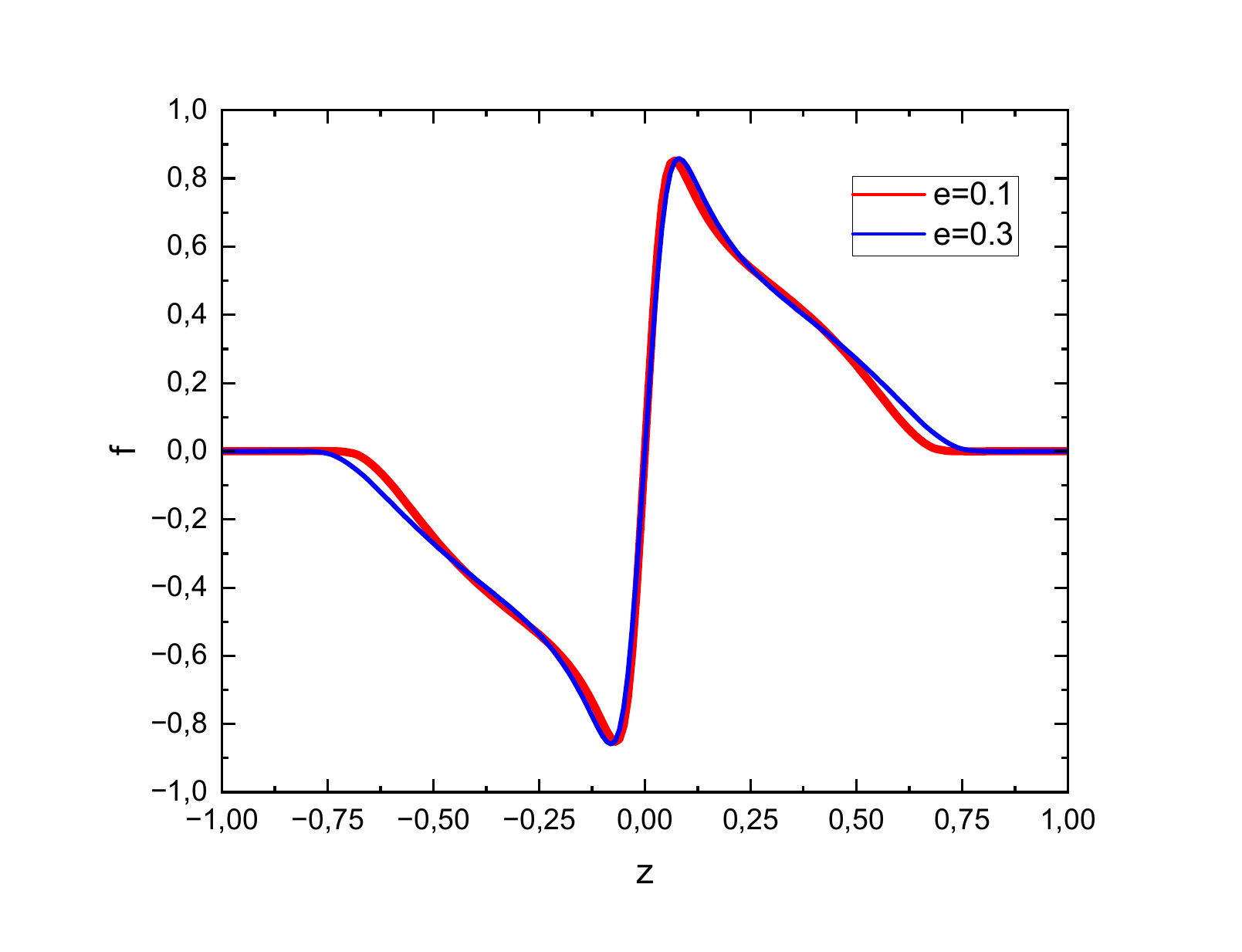}
\includegraphics[height=.28\textheight,  angle =-0]{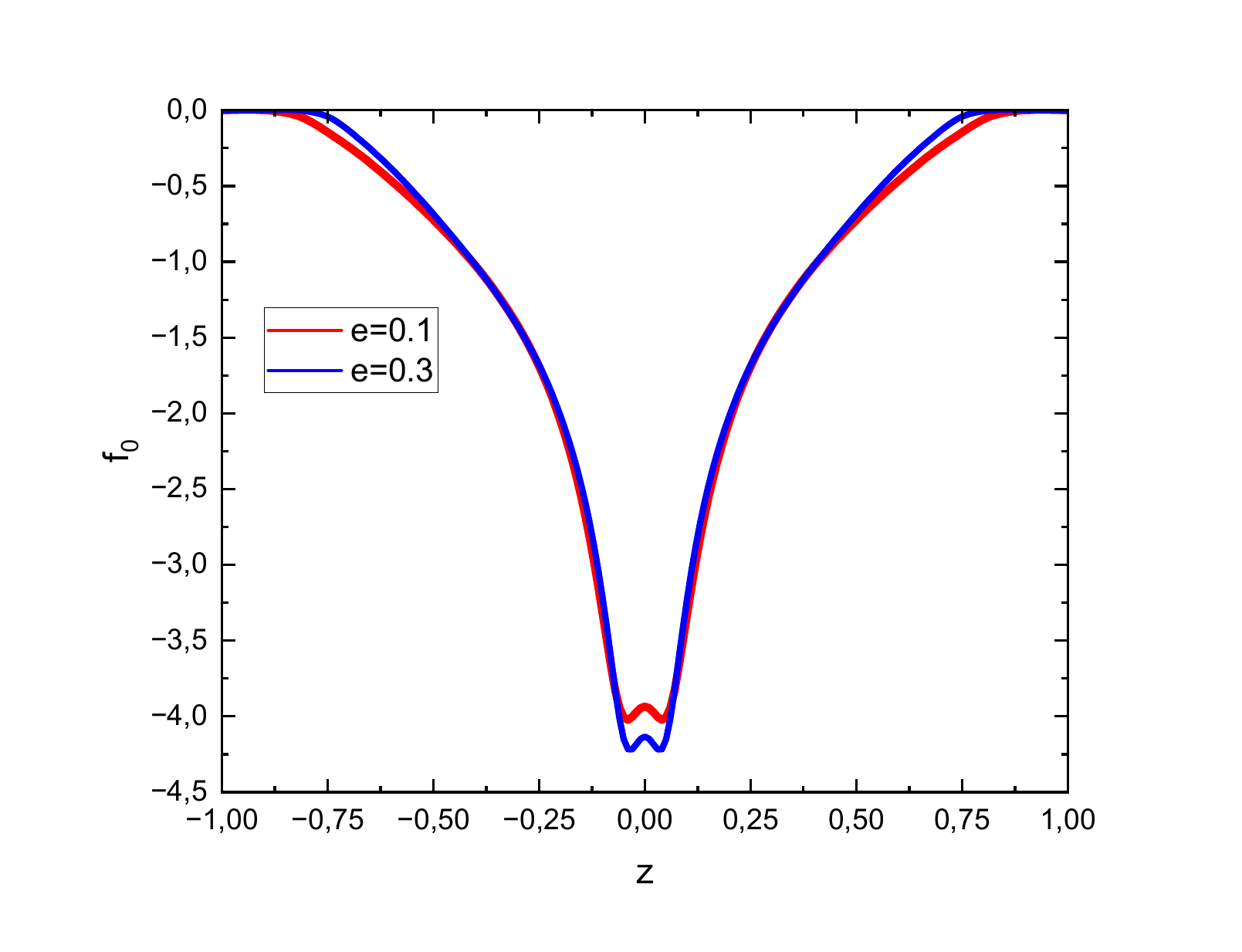}
\end{center}
\caption{\small Profiles on the symmetry axis of the $O(3)$ scalar field function (left) and the metric function $f_0$ (right) of an illustrative
$n=0$ dipole solutions  on the first branch (upper plots)  and on the third branch (lower plots) at $e=0.1$ and $e=0.3$ for 
gravitational coupling $\alpha^2=1$ and $\omega=0.885$. Here $z= r \cos \theta $ with $\theta =\frac\pi4$.
}    \lbfig{fig8}
\end{figure}
The dipole $O(3)$ configuration always possess two distinct components, located on the z-axis at\footnote{The separation $z_d$ is considered as the coordinate distance, the corresponding proper distance $L$ is defined as $$
L=2\int\limits_0^{z_d}dr e^{f_1(r,0)}
$$}  $z=\pm z_d$, see Fig.~\ref{fig8}.  

As seen in Fig.~\ref{fig5}, left plot, the mass-frequency dependence of the dipole $O(3)$ boson star looks qualitatively similar to that found for solutions of other types. When the electromagnetic interaction is weaker that the gravitational force, a fundamental branch of charged $O(3)$ dipoles  
emerge from the Newtonian limit,
where both the mass and the charge vanish, the scalar field approaches the vacuum $\phi^3=1$ and the separation parameter $z_d$ becomes infinitely large, as demonstrated in Fig.~\ref{fig11}. The gauged $O(3)$ dipole shows a spiraling behavior with an infinite number of branches toward a limiting configuration where the component of the scalar field $\phi^3$ approaches zero. 

This pattern
changes as the electrostatic repulsion becomes still stronger. The gauged $O(3)$ stars then show different 
behavior, the fundamental branch does 
not possess the Newtonian limit at the maximal frequency, see Fig.~\ref{fig11}. Then, at the mass threshold, both the ADM mass and the Noether charge of the configuration are finite, as well as the separation between the components. The fundamental branch is      
followed by a spiral.

For a given
frequency $\omega$ and the gauge coupling $e$, the mass of a dipole configuration is smaller than the mass of two spherically-symmetric gauged $O(3)$ boson stars with the same frequency. This is
consistent with interpretation of the dipole as a bound state of two individual boson stars with negative binding energy. 

We shall emphasize the important difference between the dipole of the boson stars in a theory of complex scalar field coupled to Einstein gravity \cite{Yoshida:1997nd,Herdeiro:2021mol,Cunha:2022tvk,Ildefonso:2023qty} and the corresponding self-gravitating solutions of the $O(3)$-non-linear sigma model. In the former case, the evolution along the spiral is related with an unrestricted growth of the magnitude of the scalar field while each component of the dipole becomes highly localized on the axis of symmetry. In contrast, the magnitude of the scalar field of the $O(3)$-non-linear sigma model cannot grow without limit, the spiral-like frequency dependence of the mass of the solutions towards a limiting solution is related with appearance of burls on the tail of the deformed configuration, see Fig.~\ref{fig8}. The coordinate distance $d_z$ between the two separated components decreases and the position of two individual components of the dipole, associated the two antisymmetric peaks, is clearly related to the minima of the metric function $f_0$. Unfortunately, the approach toward the limiting solution is restricted by numerical accuracy of our scheme.

\begin{figure}[h!]
\begin{center}
\includegraphics[height=.36\textheight,  angle =-0]{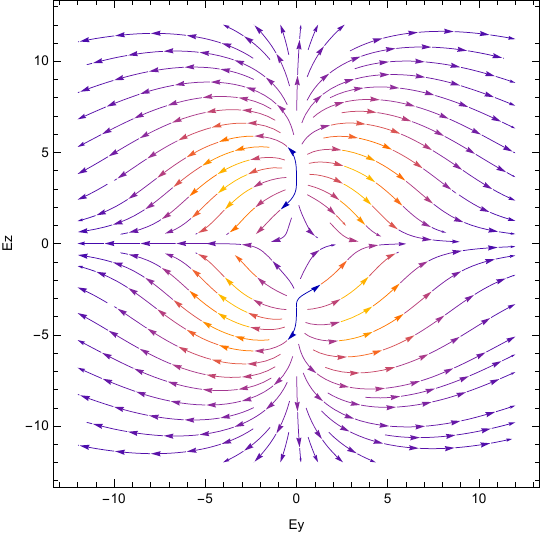}
\end{center}
\caption{\small The electric field fluxes in the $y-z$ plane of an illustrative
$n=0$ dipole solutions on the first branch for the  gauge coupling $e=0.2$ and for  
gravitational coupling $\alpha^2=1$ and $\omega=0.85$. 
}    \lbfig{fig9}
\end{figure}

\section*{Conclusions}

In this paper we have discussed globally regular self-gravitating $U(1)$ gauged solutions of the $O(3)$  non-linear sigma-model with a symmetry breaking potential. 
Configurations of four different types are considered, (i) Fundamental spherically symmetric solutions, (ii) Axially symmetric spinning solutions with non-zero angular momentum, (iii)  Parity-odd spinning solution  with an angular 
node on the symmetry plane, and (iv) 
a gravitationally bound pair of solitons,
the dipole. These configurations represent a few simple examples from an infinite tower of multipolar 
self-gravitating solutions of the $O(3)$  non-linear sigma-model, in particular, there are radially excited solutions which we do not consider here. All configurations are electrically charged with long-range Coulomb asymptotic of the electric field. The $U(1)$ gauged spinning solutions are also endowed with toroidal magnetic field with a magnetic dipole asymptotic of the gauge potential. 
These solitonic configurations do not have a regular flat spacetime limit, certain modification of the potential is necessary to support such solutions \cite{Verbin:2007fa,Ferreira:2025xey}. 

Our results indicate a close resemblance with the case of the multi-component boson stars in the massive, complex Klein-Gordon field theory. In both situations regular self-gravitating solitonic solutions exist only in a limited frequency range, the charge and mass of solutions
exhibit a spiral-like frequency dependence. In this context it is plausible that all multi-polar configurations of the boson stars reported in \cite{Herdeiro:2020kvf} may also exist in the $O(3)$ non-linear sigma-model minimally coupled to Einstein gravity.  

An essentially new feature of the $ U(1) $ gauged gravitating $O(3) $ solutions is that  the electromagnetic interaction may become stronger than gravitational attraction. Then the solutions do not possess the Newtonian limit.   

Another feature of the gauged $O(3)$ boson stars is that the amplitude of the scalar field cannot increase without a limit, the spiraling pattern is related with deformation of the tail of the configurations. 

An interesting extension of this work would be to study possible associated hairy black holes with an event horizon at the center of the configurations. 
Presence of the electromagnetic interaction allows 
to circumvent the no-hair theorem \cite{Mayo:1996mv} and construct synchronized spherically-symmetric Reissner–Nordstr{\"o}m black holes with charged $O(3)$ scalar hair by analogy with the solutions of the Einstein-Maxwell-scalar theory \cite{Herdeiro:2020xmb,Hong:2020miv,Kleihaus:2009kr}.
Further, it would be interesting to study $U(1)$
gauged dipole configurations endowed with a
black hole at the center. 

Another interesting question, which we hope to be addressing
in the near future, is to investigate the morphology of the ergo-regions of rotating and electrically charged configurations, which may lead to a
superradiant instability of boson stars \cite{Brito:2015oca,Cardoso:2007az}. 

We note, finally, that all configurations reported in this work are topologically trivial. However, 
the $O(3)$ non-linear sigma-model in 3+1 dimensional asymptotically flat space also may support topologically non-trivial Hopfion solutions. An analogy with the $U(1)$ gauged self-gravitating Skyrmion \cite{Kirichenkov:2023omy,Piette:1997ny,Radu:2005jp} suggests that the Einstein-Faddeev-Skyrme model may also support topologically trivial localized field configurations, aka pion stars \cite{Ioannidou:2006nn}. Investigation of such solutions  
is an intriguing direction to explore. 

\section*{Acknowledgements}
Y. S. would like to thank Luiz Ferreira, Carlos Herdeiro, Jutta Kunz and Eugen Radu
for enlightening discussions. He gratefully acknowledges
the support by FAPESP,
project No 2024/01704-6 and thanks the Instituto de F\'{i}sica de S\~{a}o Carlos; IFSC for kind hospitality. 

\begin{small}

\end{small}
\end{document}